\documentclass[modern]{aastex62}

\usepackage{amsmath,amssymb}
\usepackage{xspace}
\usepackage{graphicx}
\usepackage{epstopdf}
\usepackage{graphicx}
\usepackage{epsfig}
\usepackage{natbib}
\usepackage{verbatim}
\usepackage{morefloats} 

\usepackage{lineno}

\usepackage[caption=false]{subfig}
\usepackage{float}

\usepackage{CJK}

\def\beq{\begin{equation}}
\def\eeq{\end{equation}}
\def\bcm{}

\def\earthfinder{{\it EarthFinder}}

\def\mos{$\mathrm{ms}^{-1}$}
\def\micron{$\mu$m}

\usepackage{xcolor,fontawesome}
\definecolor{twitterblue}{RGB}{64,153,255}
\newcommand{\twitter}[1]{\href{https://twitter.com/#1}{\textcolor{twitterblue}{\faTwitter}\,\tt \textcolor{twitterblue}{@#1}}}

\graphicspath{{./}{figures/}}

\received{January 1, 2018}
\revised{January 7, 2018}

\submitjournal{AJ}

\newcommand{\angstrom}{\mbox{\normalfont\AA}}
 
\shorttitle{EarthFinder Tellurics}
\shortauthors{Wang et al.}

\begin{document}

\begin{CJK*}{UTF8}{gbsn}

\title{Characterizing and Mitigating the Impact of Telluric Absorption in Precise Radial Velocities}

\correspondingauthor{Sharon Xuesong Wang}
\email{sharonw@mail.tsinghua.edu.cn}

\author[0000-0002-6937-9034]{Sharon Xuesong Wang (王雪凇)}
\affiliation{Department of Astronomy, Tsinghua University, \\
Beijing 100084, China\\ 
\twitter{sharonxuesong}}

\author[0000-0001-8079-1882]{Natasha Latouf}
\affiliation{George Mason University \\
4400 University Drive \\
Fairfax, VA 22030, USA \\
\twitter{nertushka}}

\author[0000-0002-8864-1667]{Peter Plavchan}
\affiliation{George Mason University \\
4400 University Drive \\
Fairfax, VA 22030, USA \\
\twitter{PlavchanPeter}}

\author[0000-0001-6279-0595]{Bryson Cale}
\affiliation{George Mason University \\
4400 University Drive \\
Fairfax, VA 22030, USA}

\author[0000-0002-6096-1749]{Cullen Blake}
\affiliation{University of Pennsylvania \\
Philadelphia, PA 19104, USA}

\author[0000-0003-3506-5667]{\'Etienne Artigau}
\affiliation{Universit\'e de Montr\'eal, D\'epartement de Physique, IREX, Montr\'eal, QC H3C 3J7, Canada}
\affiliation{Observatoire du Mont-M\'egantic, Universit\'e de Montr\'eal, Montr\'eal, QC H3C 3J7, Canada}

\author[0000-0002-9548-1526]{Carey M. Lisse}
\affiliation{Johns Hopkins University Applied Physics Laboratory, \\
Laurel, MD USA}

\author[0000-0002-2592-9612]{Jonathan Gagn\'e}
\affiliation{Plan\'etarium Rio Tinto Alcan, Espace pour la vie, 4801 av. Pierre-De Coubertin, Montr\'eal, QC H1V~3V4, Canada}
\affiliation{Institute for Research on Exoplanets, Universit\'e de Montr\'eal, D\'epartement de Physique, C.P.~6128 Succ. Centre-ville, Montr\'eal, QC H3C~3J7, Canada} 

\author[0000-0002-1503-2852]{Jonathan Crass}
\affiliation{Department of Physics, University of Notre Dame \\
Notre Dame, IN 46556, USA \\
\twitter{jc8654}}

\author{Angelle Tanner}
\affiliation{Mississippi State University, Department of Physics and Astronomy, \\
355 Lee Blvd, Mississippi State, MS 39762, USA}

\begin{abstract}
Precise radial velocity (PRV) surveys are important for the search of Earth analogs around nearby bright stars. Such planets induce a small stellar reflex motion with RV amplitude of $\sim$10~cm/s. Detecting such a small RV signal poses important challenges to instrumentation, data analysis, and the precision of astrophysical models to mitigate stellar jitter. In this work, we investigate an important component in the PRV error budget --- the spectral contamination from the Earth's atmosphere (tellurics). We characterize the effects of telluric absorption on the RV precision and quantify its contribution to the RV budget over time and across a wavelength range of 350~nm -- 2.5\micron. We investigate the effectiveness in mitigating tellurics using simulated spectra of a solar twin star with telluric contamination over a year's worth of observations, and we extracted the RVs using two commonly adopted algorithms: dividing out a telluric model before performing cross-correlation or Forward Modeling the observed spectrum incorporating a telluric model. We assume various degrees of cleanness in removing the tellurics, including mimicing the lack of accurate knowledge of the telluric lines by using a mismatched line profile to model the ``observed" tellurics. We conclude that the RV errors caused by telluric absorption can be suppressed to close to or even below the photon-limited precision in the optical region, especially in the blue, around 1-10~cm/s. At red through near-infrared wavelengths, however, the residuals of tellurics can induce an RV error on the m/s level even under the most favorable assumptions for telluric removal, leading to significant systematic noise in the RV time series and periodograms. If the red-optical or near-infrared becomes critical in the mitigation of stellar activity, systematic errors from tellurics can be eliminated with a space mission such as \earthfinder.
\end{abstract}

\keywords{methods: numerical; techniques: radial velocities, high angular resolution}

\section{Introduction} \label{sec:intro}

Precise radial velocities (PRV) are important for finding nearby Earth analogs that are the primary targets of future direct imaging missions (e.g., \citealt{howard2016,dressing2019,morgan2021}). Recently, a number of new RV spectrographs with extremely high precision down to 10--30 cm/s have come online, nearly feasible for detecting Earth analogs around Sun-like stars \citep{eprv3}. However, using a spectrograph capable of reaching 10 cm/s does not mean achieving such a high precision on real targets, because there are still challenges in the data analyses and the mitigation of stellar jitter --- the intrinsic RV signals produced by the stars themselves, unrelated to planets \citep{exopag2015,eprv2016,halverson2016}. Mitigating stellar jitter requires a suite of efforts, including making advances in the observational and theoretical understanding of a variety of stellar astrophysics \citep{crass2021}. Similarly, reducing the error budget in the data analyses is also a multi-facet problem, including tackling many numerical and statistical challenges interwoven with observational constraints. One of the most important and urgent issues in PRV data analyses is dealing with the Earth's atmospheric contamination, considered one of the ``seven pillars holding up tent'' of the overall PRV error budget \citep{eprv2016}. 

The Earth's atmosphere imprints numerous absorption and emission lines (hereafter telluric lines) on top of an emitted stellar spectrum, and these lines do not exhibit the same Doppler shift that the stellar lines do due to the Earth's spin and its motion around the Sun. Therefore, the blending between the telluric lines and the stellar lines would lead to errors when estimating the stellar Doppler shift. Telluric contamination has been a well known bottleneck for achieving higher RV precision in the near infra-red (NIR; \citealt{bean2010}) at the meter per second (m/s) level. However, in the optical, which is relatively free of telluric lines, there are in fact prevalent shallow water absorption lines, called the ``micro-telluric" lines, and multiple bands of oxygen lines. These micro-telluric lines (flux depths $<$2 \% and mostly $<$1 \%) at visible wavelengths can contribute to the RV error budget at the $\sim$20 cm/s level \citep{cunha2014,artigau2014,lisogorskyi2019}.

Eliminating the impact of telluric lines in PRV can be challenging for multiple reasons. While emission lines can, in principle, be removed via sky subtraction, it is still very challenging to achieve a thorough removal due to the variability of these emission lines and th einherent noise in low SNR sky spectra obtained using a sky calibration fiber (e.g., low SNR). The only exception is when using spectrographs fed with diffraction-limited images through single-mode fibers, which would have a significant reduction in sky emission lines (as well as moon contamination; \citealt{Crepp2016}) thanks to its small point-spread function. More challenging yet are the absorption lines, especially in the NIR, where deep lines are common even within the few spectral windows in between the saturated lines. Early work often took observations of telluric standard stars (hot stars with relatively flat continuum) to characterize and subtract the telluric absorption lines \citep{vacca2003}, and the best effort so far in this front has achieved a residual around 2\% for water lines in the 0.9--1.35\micron\ window \citep{sameshima2018}, but often around 5\% or so elsewhere. However, they note that the time and spatial variability of water absorption puts stringent requirements on the telluric standard star observations. 

There are several packages that could generate synthetic telluric absorption lines to fit the observations, e.g., TAPAS \citep{tapas}, Telfit \citep{gullikson2014}, MolecFit \citep{smette2015}, and other packages based on HITRAN \citep{hitran2013,hitran2017}. The best-effort in modeling the observed spectra using synthetic lines are relatively successful, with a residual level of around 1--2\%, for example, in fitting a high-resolution, high-fidelity solar spectrum \citep{baker2020}. \citet{ulmer-moll2019} compared several methods and also most commonly used software packages to correct for telluric absorption lines using synthetic spectra, and their best results typically have residuals around 3-7\% (1-2\% at the very best), on a similar level with \cite{sameshima2018}. Modeling residuals at this level could lead to an RV error on the order of 20--50~cm/s or even more \citep{sithajan2016}, similar to the impact of micro-tellurics in the optical region.

There are multiple challenges in improving the modeling precision of telluric absorption lines. Water lines, copious and highly variable in time (though less so along lines of sight; \cite{li2018}), are the most challenging lines to model. It is difficult to know the amount of precipitible water vapor (PWV) in the atmosphere during an observation a priori, which is an important model parameter governing the line depths. Although there are ways to measure PWV though remote sensing data from satellites \citep{meier2021} and using GPS \citep{baker2017}, these methods are not yet precise enough to a sub-mm level. Other challenges with modeling the absorption lines, especially the relatively deeper ones, include imperfections in the line lists and line profiles \citep{seifahrt2010,baker2020}, as well as Doppler shifts of the lines due to the wind \citep{figueira2012}.

Overall, telluric contamination is a large term in the PRV error budget that is challenging to mitigate, but this is also a term that can be eliminated completely by going to space, which is one of the motivations behind the NASA Probe Mission Concept Study, \textit{EarthFinder} \citep{plavchan2020}. As part of the \textit{EarthFinder} study, our work set out to investigate the gain in terms of the RV precision improvement when eliminating telluric contamination completely by going to space. To be specific, our goals are:\footnote{We limit the scope of our work to telluric absorption, as it is significantly more challenging to deal with than the emission lines.} (1) to quantify the RV precision limit set by the telluric contamination in a broad wavelength range, from 350 nm to 2.5 μm, on a Sun-like star; (2) to characterize the effectiveness of commonly used methods for mitigating tellurics (i.e., dividing out the telluric lines and Forward Modeling). We perform simulations with synthetic spectra and extract RVs from these spectra to assess the RV precision limit set by the telluric contamination for ground-based instruments. With our controlled simulations, we isolate the effects of tellurics from other factors such as photon noise, stellar activity/jitter, and instrumental effects.

The paper is structured as follows. In Section~\ref{sec:method} we describe our methods in synthesizing the observed spectral time series, extracting radial velocities while mitigating telluric lines, and combining RVs across different wavelengths. Section~\ref{sec:results} describes our main results and the interpretation, including how tellurics enter the RV error budget across the wavelengths, what correlates with telluric induced errors, the net contribution of tellurics to the RV error budget in different photometric bands, and how tellurics affect the RV time series measurements. Section~\ref{sec:discussion} enumerates the caveats in this study, summarizes our major conclusions, and outlines a few aspects of important future directions.

\section{Methodology} \label{sec:method}

\begin{figure}[!tb]
    \centering
\includegraphics[trim=0.3cm 0cm 0cm 0cm,clip,scale=.55]{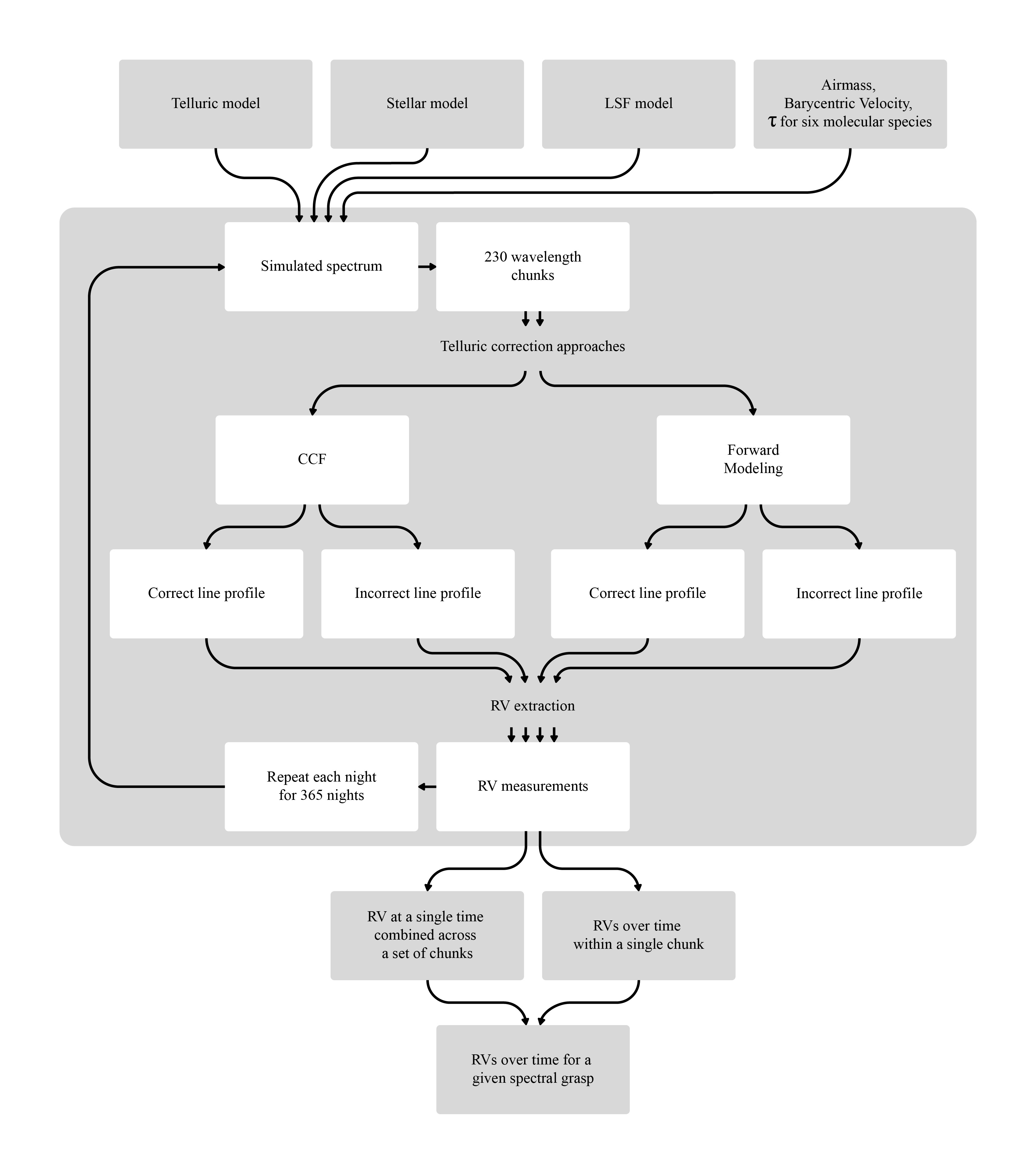}
\caption{Flowchart illustrating our simulation methodology, including the generation of synthetic spectra and the RV extraction process using the Cross-Correlation Function (CCF) and the Forward Modeling (FM) methods with both correct and incorrect line profiles. The process of RV extraction is repeated each night for 365 nights for each method and line profile, and then they are combined to form a single nightly RV.
\label{fig:flowchart}}
\end{figure}

Overall, our simulation aims to reproduce an ensemble of spectra of a Sun-like star through the Earth’s atmosphere throughout a year. We then extract RVs from these spectra using different telluric mitigation methods, and combine the RVs from different wavelengths through a weighted-average scheme. We then compare the RVs extracted using different methods and telluric mitigation strategies and compare their effectiveness. Figure~\ref{fig:flowchart} illustrates the workflow of our work.

\subsection{Simulating a Year of Observed Spectra}

We simulated a set of 365 spectra of a Sun-like star, one per night over a year. There are two spectral components in each of our simulated spectrum: the stellar spectrum and the telluric spectrum. We use the Kurucz solar spectrum generated by ATLAS9 \citep{kurucz2005}, and we keep the solar spectrum constant throughout our simulation, i.e., no stellar line variation due to stellar jitter. We multiply telluric absorption spectrum onto the solar spectrum using synthetic telluric spectra generated by TAPAS \citep{tapas}. The telluric spectrum is different for each simulated night, as described below.

There are two time-variable components in our simulated spectra:

\textbf{(1) The telluric spectrum: }We varied the depths of the telluric lines according to three things: the atmospheric temperature and pressure profiles of the observing season, the airmass of each observation, and the amount of PWV for each simulated night. We assumed the observations were taken on a telescope on Kitt Peak and adopted the corresponding atmospheric profiles. To mimic the change caused by atmospheric temperature and pressure profiles in the telluric spectra, we generated the telluric spectra using TAPAS for a selected date of each of the twelve months throughout the year of 2017 (usually the 15th of each month, unless the atmospheric profile data are unavailable). To be specific, for each month, we generated a suite of spectra for water, methane, carbon dioxide, nitrous oxide, oxygen, and ozone molecules separately, without additional line broadening or Rayleigh scattering, and at the highest resolution and sampling possible, at an airmass of 1.0. We note that only deep lines would exhibit visible differences across the season, presumably due to a change in the atmospheric conditions that TAPAS retrieves.

\begin{figure}[ht]
\includegraphics[scale=0.43]{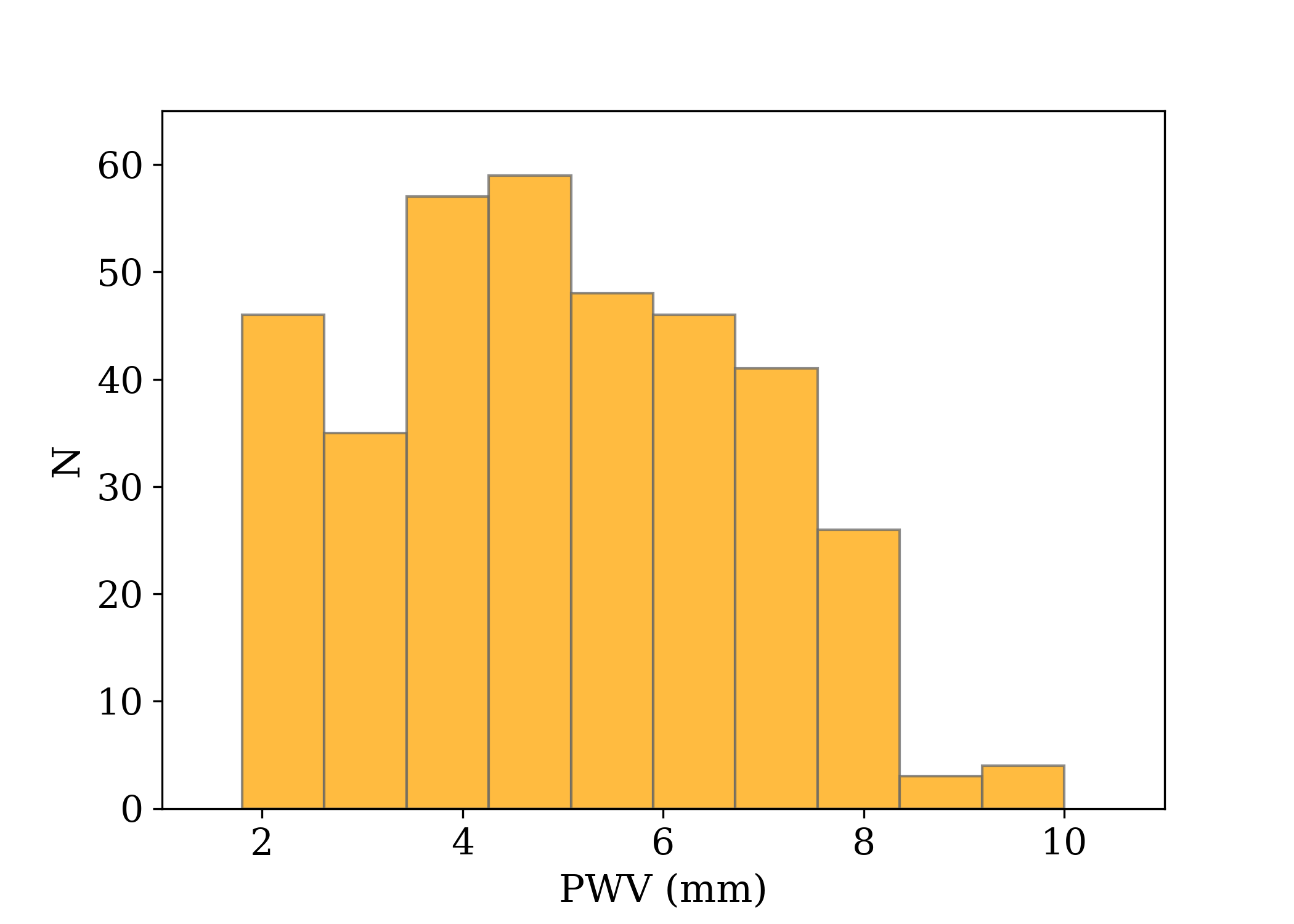}
\includegraphics[scale=0.43]{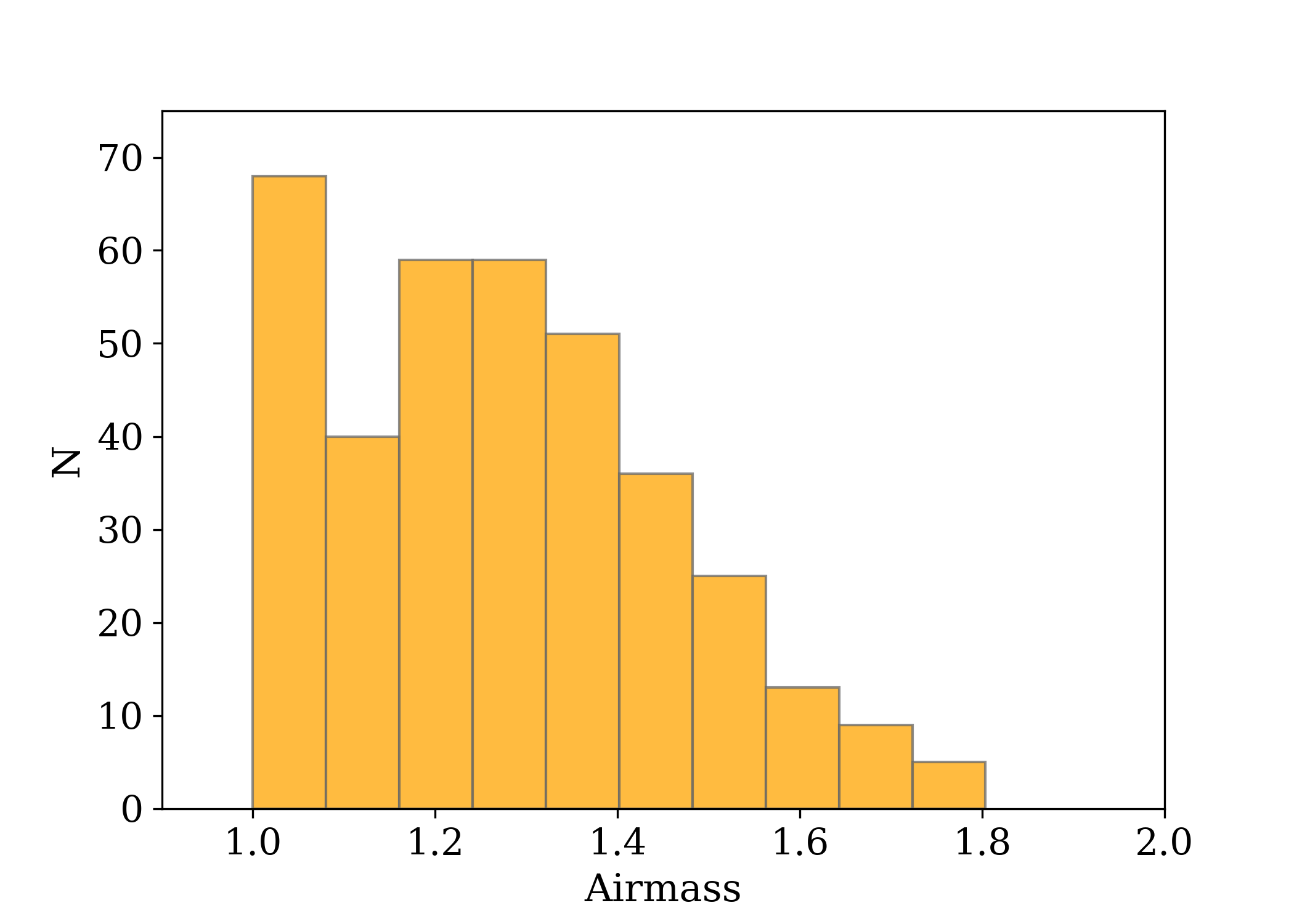}
\caption{These histograms show the distribution of the precipitable water vapor (PWV) and airmass used for the 365 nights in our simulation. The PWV values are drawn from an ensemble of real PWV measurements for Kitt Peak but restricted between 2~mm and 10~mm, and the airmass values are drawn from real Keck/HIRES observations on RV standard stars.
\label{fig:pwvhist}}
\end{figure}

In addition, we randomly drew a value for the amount of PWV from a Gaussian distribution centered at 5.0mm with width 2.0mm and scale the water lines accordingly. The PWV values were drawn from a distribution that resembles the PWV values of astronomical nights at Kitt Peak in 2017 as recorded by Suominet (https://www.suominet.ucar.edu/; e.g., Wang et al. 2019).
We then combined the spectra of all six molecules through multiplication, and then scaled the product telluric spectrum by the airmass of the corresponding night. For each night, we randomly drew a number for the airmass from a truncated Gaussian distribution centered at 1.25 with width 0.2 and forced to be $>1$, and we scaled the telluric absorption to the appropriate depths according to this airmass value. The distribution of the airmass resembles the RV standard star HD 10700, or Tau Ceti, as observed by Keck/HIRES (e.g., \citealt{butler2017}). Figure~\ref{fig:pwvhist} shows the histograms of our input PWV and airmass values. 

\textbf{(2) The barycentric velocity shift: } We determined the barycentric velocity correction (BC; the relative velocity between the observatory and the star due to Earth’s motion around the solar system barycenter) of the “observed star” for that night following
\beq
\sin{\big( {\rm date} /(2\pi \cdot 365.15)-\pi/2 \big)} \cdot 30~{\rm km/s},
\eeq
which roughly matches with the BC values throughout a year for tau Ceti as observed at Keck. Then we shifted the telluric spectra according to the BC value to offset the telluric spectra from the stellar spectrum in RV space to mimic the Earth’s motion. We chose to shift the telluric spectrum and leave the stellar spectrum untouched, because the TAPAS synthetic spectra have a higher fidelity than the Kurucz solar spectrum, and thus this would minimize the numerical errors in shifting (interpolating and re-sampling).

Finally, we convolved the spectrum to a spectral resolution of R=120,000, which is representative of the spectral resolution of the next-generation ground-based PRV instruments. We did not bin the spectrum further down to mimic the CCD pixels but chose to preserve the original binning of the Kurucz solar spectrum to avoid further numerical errors. As a result, the “pixel size” in our simulated spectra is about 0.0007-0.005 nm, a factor of three smaller than the typical CCDs used for PRV instruments today. We chose not to add photon noise or blaze functions in our simulated spectra to avoid adding more RV errors in order to isolate the effects of tellurics. Then, we extracted the RVs through two methods, using the cross-correlation method or Forward Modeling.

\subsection{Extracting RVs and Correcting for the Telluric Absorption}

We extracted the RVs from our simulated spectra with two methods that are commonly employed in the community: the cross-correlation function (CCF) method, where RVs are estimated from the CCF between the observed and a template spectrum (or a mask), or the Forward Modeling method, where the observed spectra are modeled with the input spectral templates and several free parameters, including the Doppler shift. In each method, we used a perfect stellar template that is identical to our input stellar spectrum (i.e., the synthetic solar spectrum) to minimize additional errors unrelated to the tellurics. Whenever needed, we assumed perfect knowledge of the wavelength solution on the observed spectra and the spectral point spread function (PSF; also called the line spread function or the spectrograph response function), an assumption not too far from reality for the latest generation of ultra-stabilized RV spectrographs calibrated by laser frequency combs. With each method, we apply a set of telluric mitigating strategies with various degrees of accuracy in eliminating tellurics. 

Before extracting the RVs, we divided each spectrum into 230 chunks (``orders"), from 0.3 nm at 350 nm to $\sim$20 nm at 2.5 $\mu$m, with the chunk size growing linearly versus wavelength. The change of the chunk size is to mimic the length of a typical spectral order and also to compensate for the decrease of Doppler content towards the NIR. Computing RVs in the unit of spectra orders is also convenient for characterizing the RV errors induced by tellurics as a function of wavelength. This is also close to reality where RVs are typically computed independently for different spectral orders, and then combined together later on.

\begin{figure}[!htb]
\includegraphics[scale=1.0]{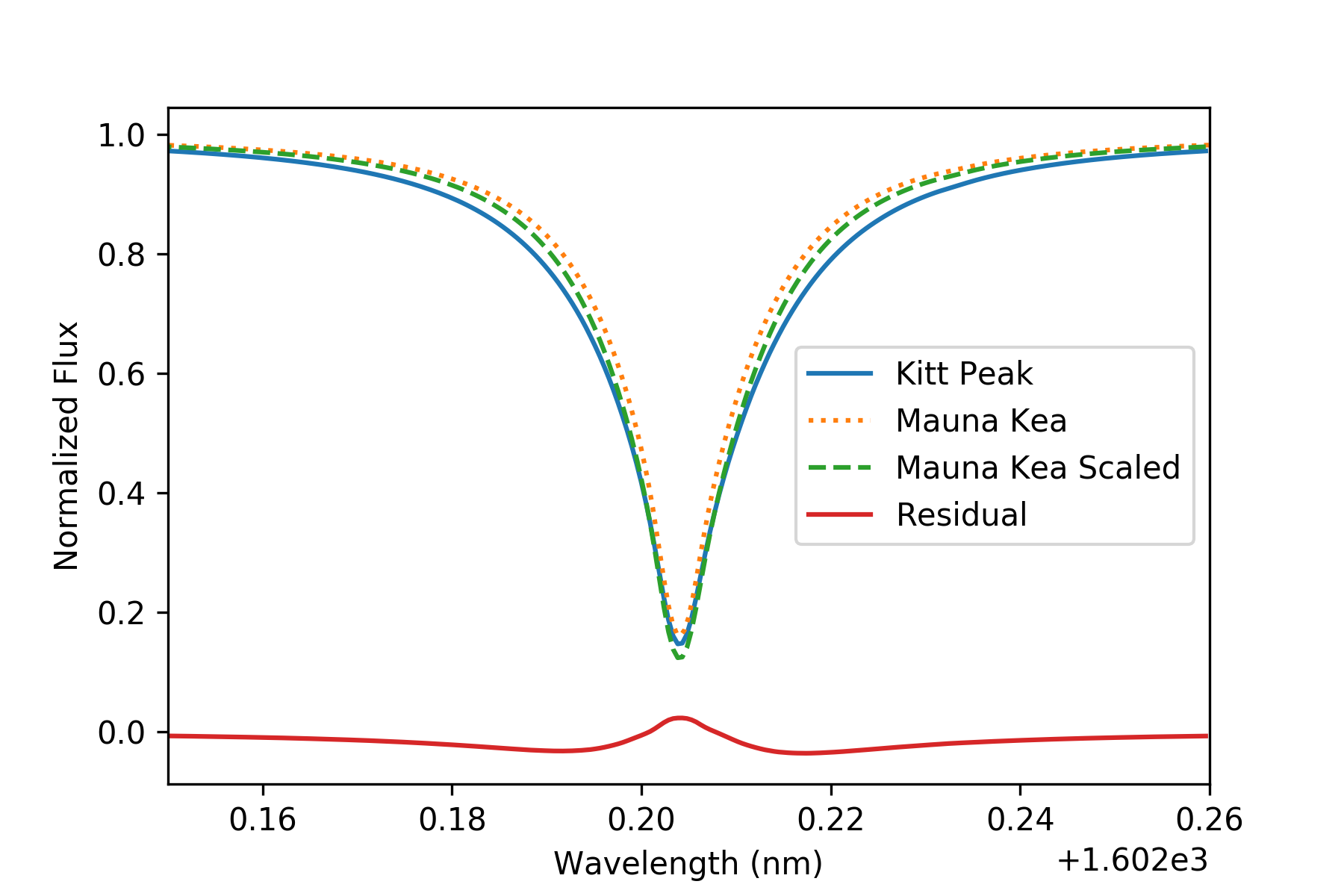}
\caption{Comparison of the line profiles for a CO2 line near 1602 nm for two observatories: Kitt Peak (used in simulated spectra) and Mauna Kea (used as input to fit the simulated spectra). Both spectra are at airmass = 1. The Mauna Kea Scaled spectrum is the best-fit version when fitting the Kitt Peak profile by scaling the Mauna Kea line by a power law. The RMS of the residual of this fit is 1.2\%, which is around the typical value for all lines and smaller than the typical residuals (2--5\%) reported by \citet{ulmer-moll2019}.
\label{fig:lineprofile}}
\end{figure}

We describe each method and the associated mitigation strategy in detail below:
\begin{enumerate}
\item  No correction: RVs are extracted via the CCF method as first described by \citet{baranne1996} but using a stellar template instead of a mask and stepping through the RV space instead of fitting a gaussian to the CCF, similar to \cite{zeichmeister2018}.\footnote{We had performed a sanity check on our CCF algorithm to ensure that an RV of 0~m/s are returned from all orders when extracting RVs from a simulated solar spectrum with no tellurics. This confirms that we do not introduce numerical errors beyond the machine precision or due to any edge effects in our choice of chunking up the spectra or in our CCF algorithm.} We do not perform any correction on the tellurics, so this method characterizes the net contribution of tellurics to the RV error budget when left untreated. We note that using a Forward Modeling method gives the same results, so we simply present the results from the CCF method here for the ``No correction" method.

\item CCF after telluric division (referred to as ``CCF with division" or ``Division" below): RVs are extracted via the CCF method, but before the CCF is computed, a telluric spectrum is removed from each order via mathematical division. The telluric spectrum being divided out is exactly the same as the one used for generating the simulated spectrum for each night, but convolved to R=120,000 to match the resolution of the observed spectrum. This approach is the most commonly used method to mitigate tellurics in PRVs for stabilized spectrographs such as HARPS and SPIRou \citep{cretignier2021,artigau2021}.

\item CCF after telluric division, but with an incorrect telluric line profile (``CCF with wrong profiles" or ``Division K-profile"): RVs are extracted in the same way as the CCF with division method described above. However, instead of using the exact same telluric spectrum as the input, we use a spectrum with a different line profile setting than the input telluric spectrum. Our input telluric spectrum was generated assuming the atmospheric conditions over Kitt Peak, and here in this method, we divide out a spectrum generated using conditions over Mauna Kea and hence differing in the underlying line profiles. We scale the Mauna Kea telluric spectrum to match the line depths in the simulated observed spectrum to mimic a modeling process,\footnote{This is equivalent as knowing the line depths to some degree (though not perfectly) prior to the final RV extraction, a fairly common situation in reality (e.g., \citealt{artigau2021}).}, with the exception of ozone absorption, which we omit completely from the division, because the ozone absorption is more like a modulation on the continuum level and is very hard to know a priori. A comparison of the two line profiles is shown for an example telluric line in Figure~\ref{fig:lineprofile}. The difference in the two sets of lines are mostly due to atmospheric pressure difference effects between the two sites. Introducing such a difference is also similar to having the line width wrong. This introduction of a line profile mismatch is to mimic our lack of knowledge on the telluric lines, similarly to having a modeling residual of around 1\% for typical lines (somewhat smaller than the best effort in the litarature, e.g., \citealt{ulmer-moll2019}).

\item Forward Modeling with the perfect input telluric spectrum: RVs are extracted via the Forward Modeling method as described in \citet{butler1996}. This method builds a model using the stellar template and the synthetic telluric spectra (both the same as used in simulation input), with the only free parameter being the Doppler shift (RV), to fit the observed spectrum via a least-square minimization algorithm.\footnote{We use the \texttt{LMFIT} package in python with the L-BFGS algorithm, which we observed to have the best convergence properties for our calculations.} We again assume perfect knowledge on the wavelength solution and the spectral PSF. This method serves as the sanity check for our work, and as expected, it returns 0~m/s (to a precision of $10^{-13}$ m/s) in all orders across all nights, meaning no RV errors. We do not show the results from this method for the rest of the paper.

\item Forward Modeling with an incorrect telluric line profile (``modeling with wrong profiles" or ``modeling K-profile"): RVs are extracted following the same methodology as described above, however we again introduce a different telluric line profile than the one used to generate the synthetic spectra. Relatively similar to the CCF with wrong profiles method, we first estimate a set of best-fit line depths for the six molecules in our telluric template spectra (with the wrong line profile) by performing a first iteration Forward Modeling fit to the observed spectrum. In this first iteration, the telluric line depth for each molecular species is a free parameter in the modeling, but only for spectral orders where the Doppler information content in the lines of this particular molecular species exceeds 1~m/s (estimated following \citealt{butler1996} assuming SNR$=1,000$ per pixel). As in ``CCF with wrong profiles'', we ignore the ozone completely. If the lines do not carry enough Doppler information (e.g., too shallow or too few), we freeze the line depth to zero and do not attempt to retrieve the line depth for this species from this particular order. Next, all the best-fit line depths are averaged to report a final depth for each species, which are then fixed for all orders in the second (and final) iteration of Forward Modeling. We adopt this two-step modeling, because we find a significant improvement in the RV precision by adding a second iteration.

\end{enumerate}

\subsection{Combining RVs from Different Orders}

After RV extraction, we have a matrix of RVs across 230 orders and 365 nights. We then combine the RVs of the 230 orders from a single night into one reported RV value for the night. The combination is done through a weighted average, and the weight for each order consists of two components added in quadrature: the RV scatter across 365 nights for this order (caused by tellurics) and the photon-limited RV error. We bring back photon errors here so that the weights are more realistic, because otherwise the spectral orders with tellurics will be down weighted significantly, eliminating errors induced by tellurics almost completely. 
The RV scatter, caused by tellurics, is simply estimated by computing the standard deviation of the RVs of each order across 365 night. The photon-limited RV error is estimated following \cite{butler1996} assuming SNR$=100$ per pixel, similar to the typical SNR achieved per resolution element in typical real RV observations. At this SNR, the photon limited RV error for an order is typically 1--10s~m/s, and about half of the orders have the photon limited RV error as the dominant term in their weights. Such an weighted averaging scheme represents the typical strategy in combining RVs across multiple orders.

\section{Results and Interpretation}\label{sec:results}

\begin{figure}
\centering
\includegraphics[scale=0.5]{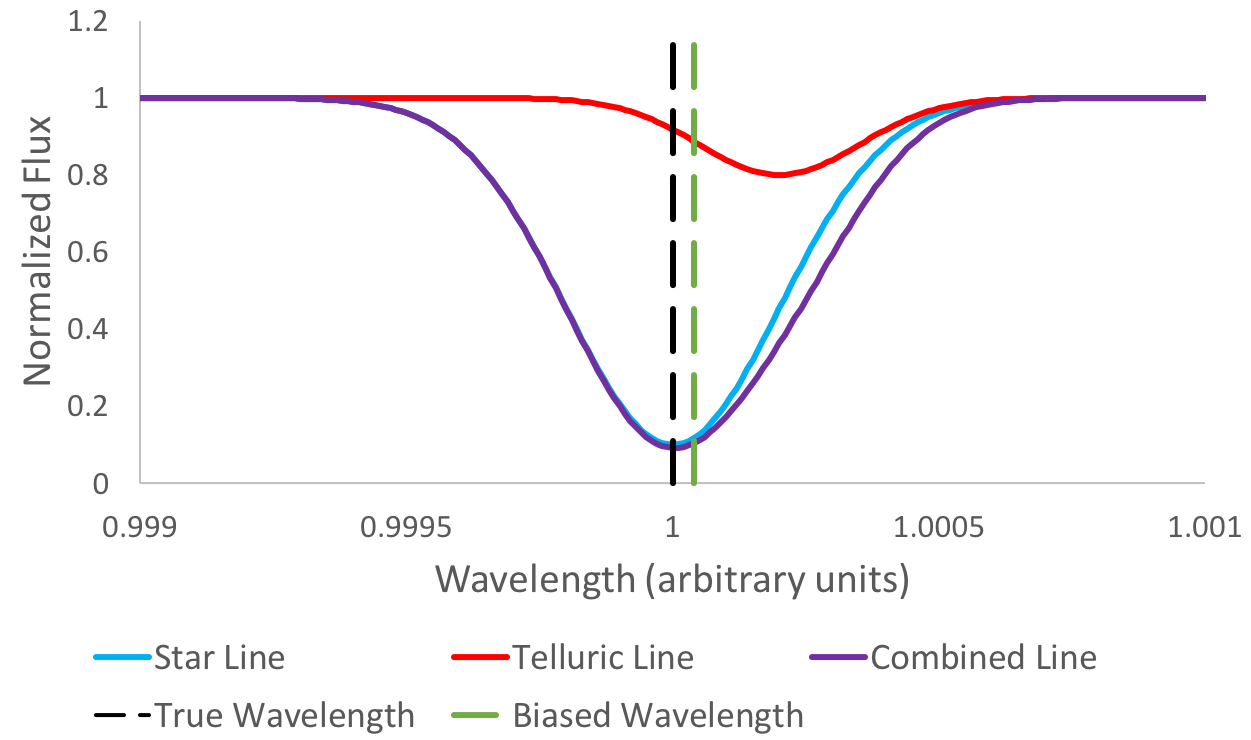}
\caption{Qualitative illustration of how the telluric lines “pull” the centroid of stellar absorption lines off from their true Doppler-shifted stellar absorption lines and introduce bias to the final measured RV \citep{wright2013}.
\label{fig:teldemo}}
\end{figure}

Overall, there are two ways that tellurics contribute to the RV error budget. One is through peak pulling (e.g., \citealt{wright2013}), where the telluric contamination lines would ``pull" the stellar spectral lines when performing CCF or modeling, as illustrated in Figure~\ref{fig:teldemo}. In other words, the existence of telluric line changes the slope of the stellar line, which is the origin of the Doppler information in a line, and this leads to a biased measurement of the RVs. 
Any residuals in dividing or modeling the tellurics would cause RV error in a similar fashion. This is very analogous to how the sky/solar contamination affects RV precision \citep{roy2020}.

\begin{figure}
\centering
\includegraphics[scale=0.7]{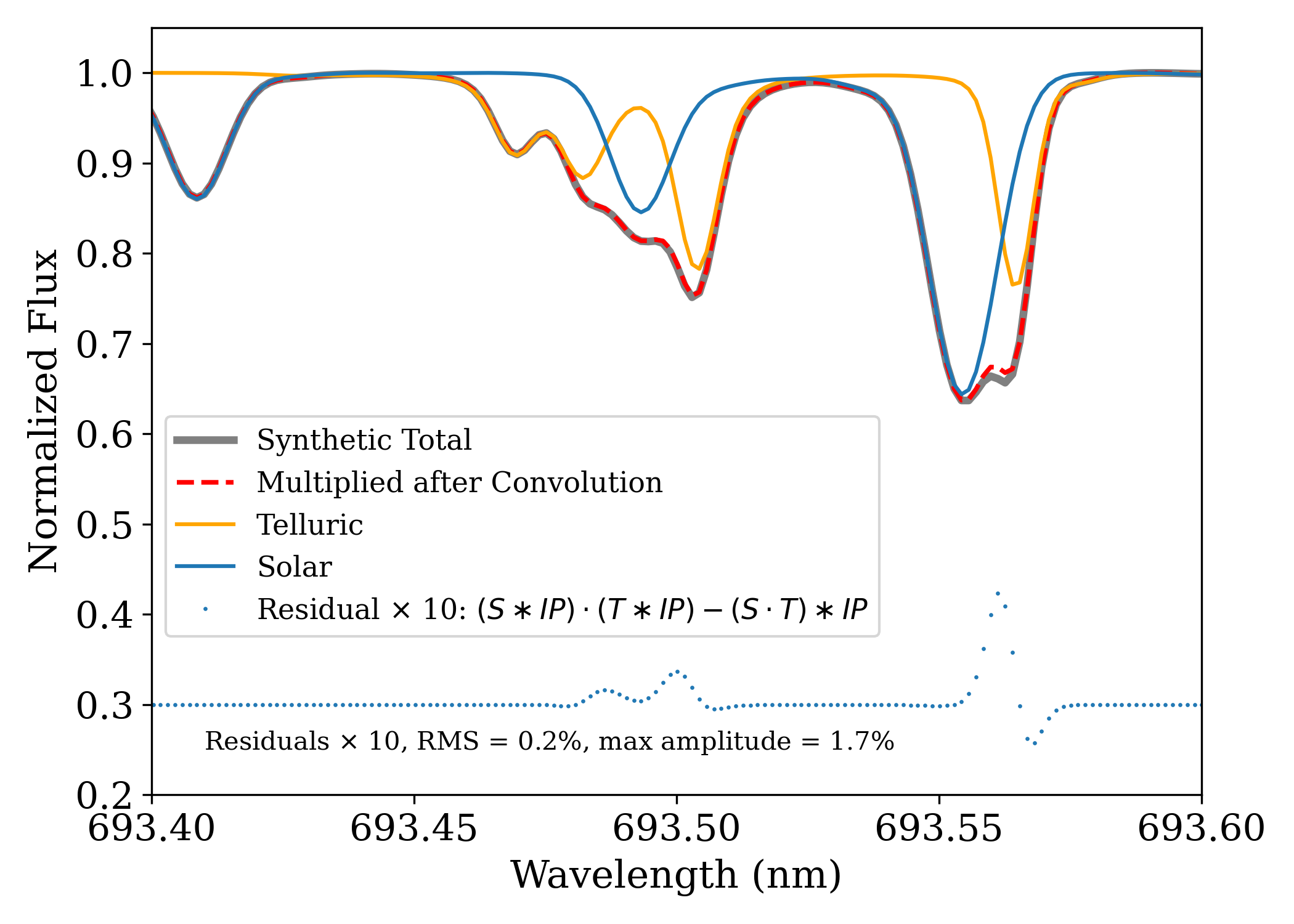}
\caption{Demonstration of how convolution does not distribute over multiplication (difference between the gray and red lines). The residuals are amplified by a factor of 10 and shifted up arbitrarily for clarity. As shown here, dividing a telluric spectrum (convolved with the spectral PSF, "IP") from an observed stellar×telluric spec-trum will introduce errors. Such an division, commonly used as a treatment for tellurics when performing CCF to extract RVs (e.g., \citealt{artigau2014}), will not result in a totally clean stellar spectrum at the \% level versus convolution with the spectral PSF.
\label{fig:conv}}
\end{figure}

The second way is more subtle, which only occurs when dividing out the telluric lines. One of the common treatment for tellurics when performing CCF to extract RVs is to divide out a telluric model, just like what we do in our simulations (e.g., \citealt{artigau2014}). However, this is mathematically incorrect, because the stellar$+$telluric spectrum is a convolution between the spectral PSF ($IP$) and a product between the \textit{unconvolved} stellar spectrum, $S$, and the \textit{unconvolved} telluric spectrum, $T$ --- i.e., multiplication first and then convolution, or the observed spectrum can be expressed as $(S\cdot T)\ast IP$. Mathematically, convolution does not distribute over multiplication, or $(S\cdot T)\ast IP \neq (S\ast IP)\cdot(T \ast IP)$, as illustrated in Figure~\ref{fig:conv}. Therefore, dividing a telluric model ($T \ast IP$) will not result in a clean stellar spectrum ($S\ast IP$). Using division to remove telluric contamination, even if using a perfectly accurate telluric model, will thus result in additional errors in the spectrum, which would then translate into RV errors. Such error is larger for deeper lines, and as a result, this effect is more severe for the NIR region. This division error does not exist in the Forward Modeling results as the Forward Modeling algorithm follows the correct order of multiplication then convolution when constructing models to fit the observed spectra. We note that since our sampling scale is smaller than the typical CCD pixel scales in real instruments, these convolution$+$division errors might be underestimated. This error also increase with a lower spectral resolution.

In the following subsections, we first take a detailed look into how tellurics affect the RVs as a function of wavelength, and then we take a more holistic view at the tellurics' overall contribution to the RV error budget under different scenarios.

\subsection{Errors Induced By Tellurics vs.\ Wavelength}

\begin{figure}[!htb]
\includegraphics[scale=0.43]{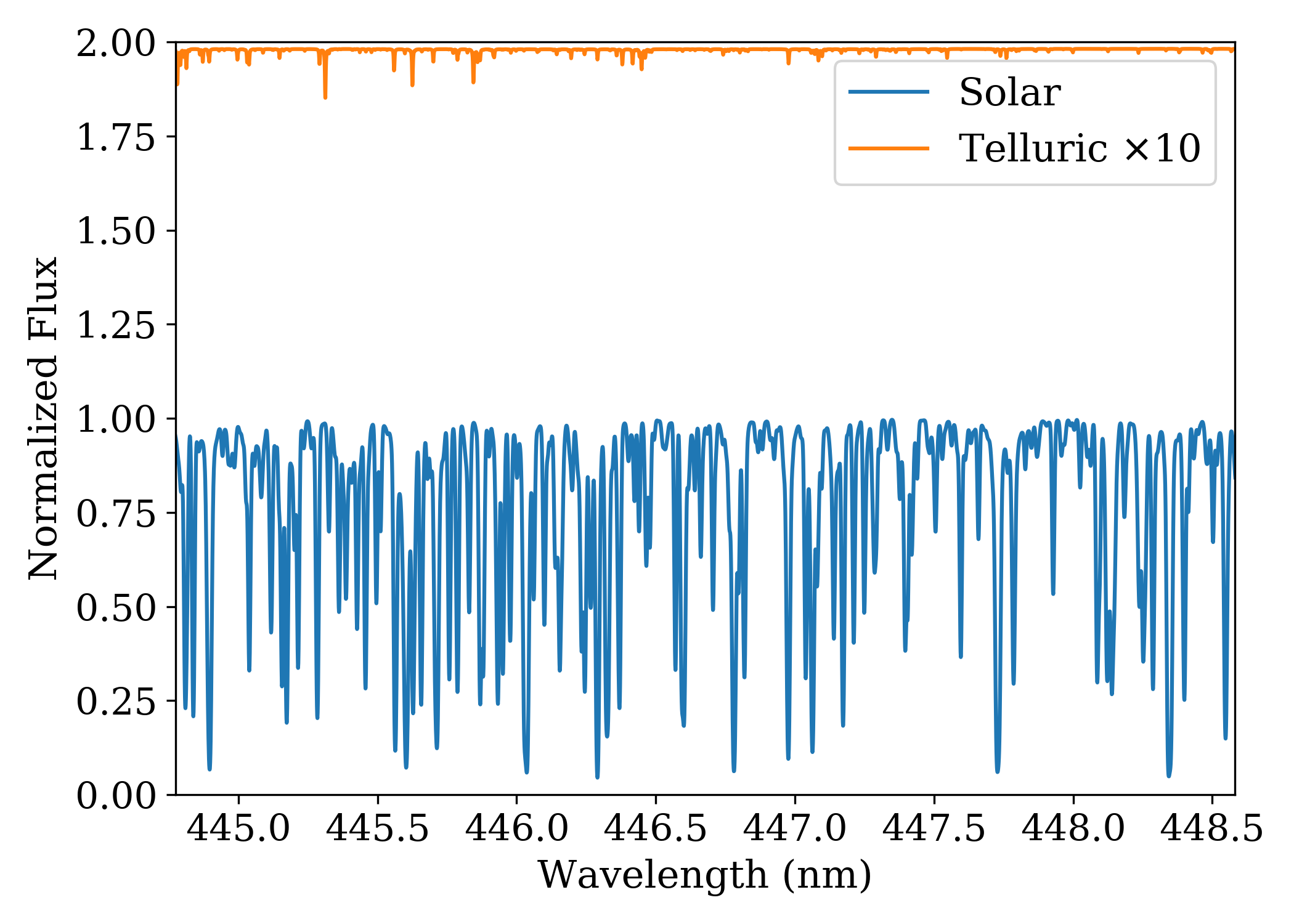}
\includegraphics[scale=0.43]{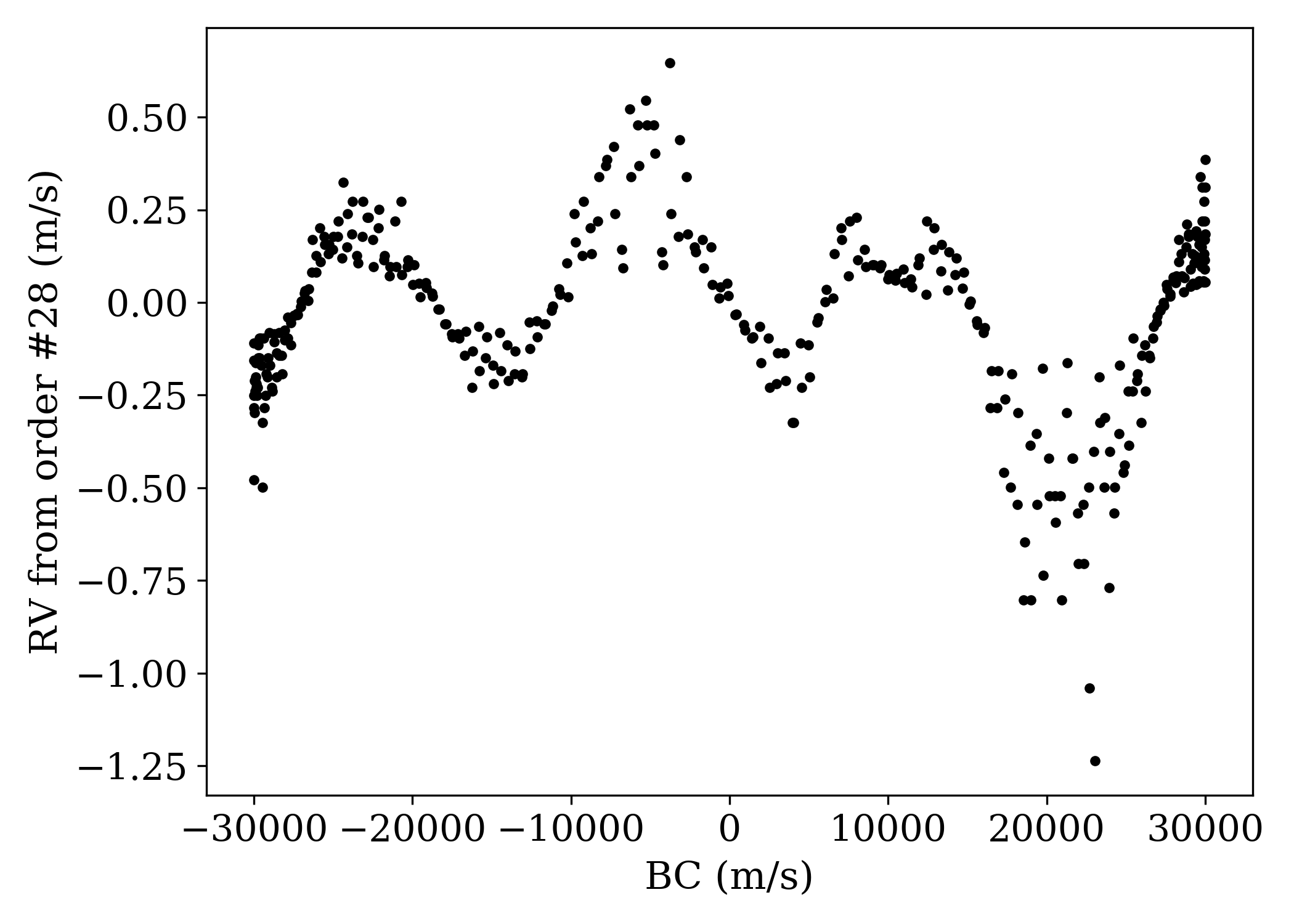} \\
\includegraphics[scale=0.43]{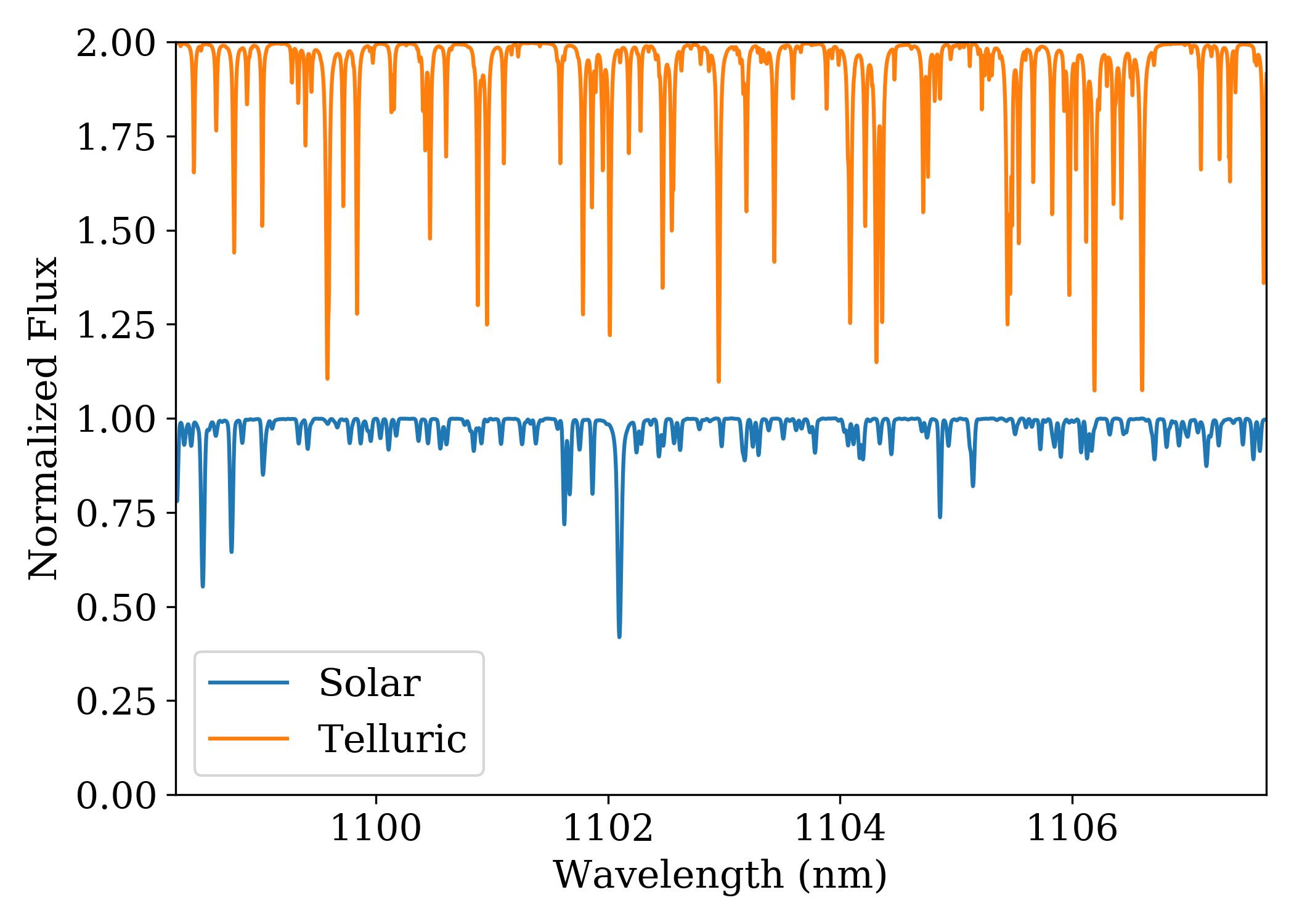}
\includegraphics[scale=0.43]{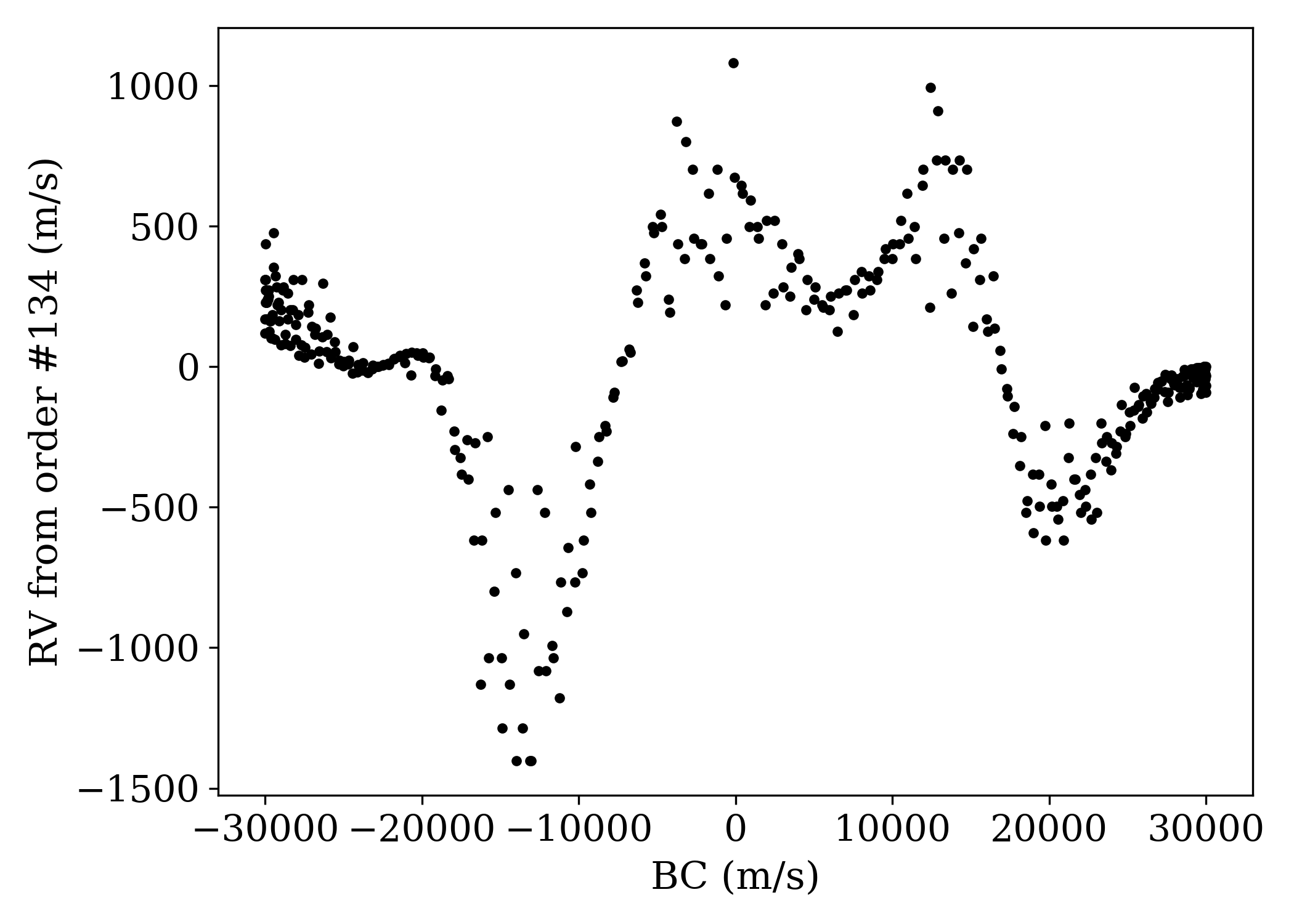}
\caption{Example spectral orders and their corresponding RVs vs.\ the barycentric correction (BC; relative velocity to the star due to the Earth's barycentric motion) when tellurics are ignored in the RV extraction. The plots on the left are the solar spectrum (blue) and the telluric spectrum (orange) for each order, with the telluric absorption on for the top plot (order \#28) amplified by 10 times for clarity. The plots on the right illustrate how telluric lines of different depths affect the RVs to a different degree, and the RV biases caused by tellurics are highly structured in the temporal space due to the Earth's barycentric motion. The RV errors in the bottom right NIR plot are a thousand times larger than the errors in the upper right blue-optical plot because the solar lines are much weaker and the telluric lines are much stronger in the NIR.
\label{fig:chunks}}
\end{figure}

\begin{figure}[!htb]
\centering
\includegraphics[scale=0.4]{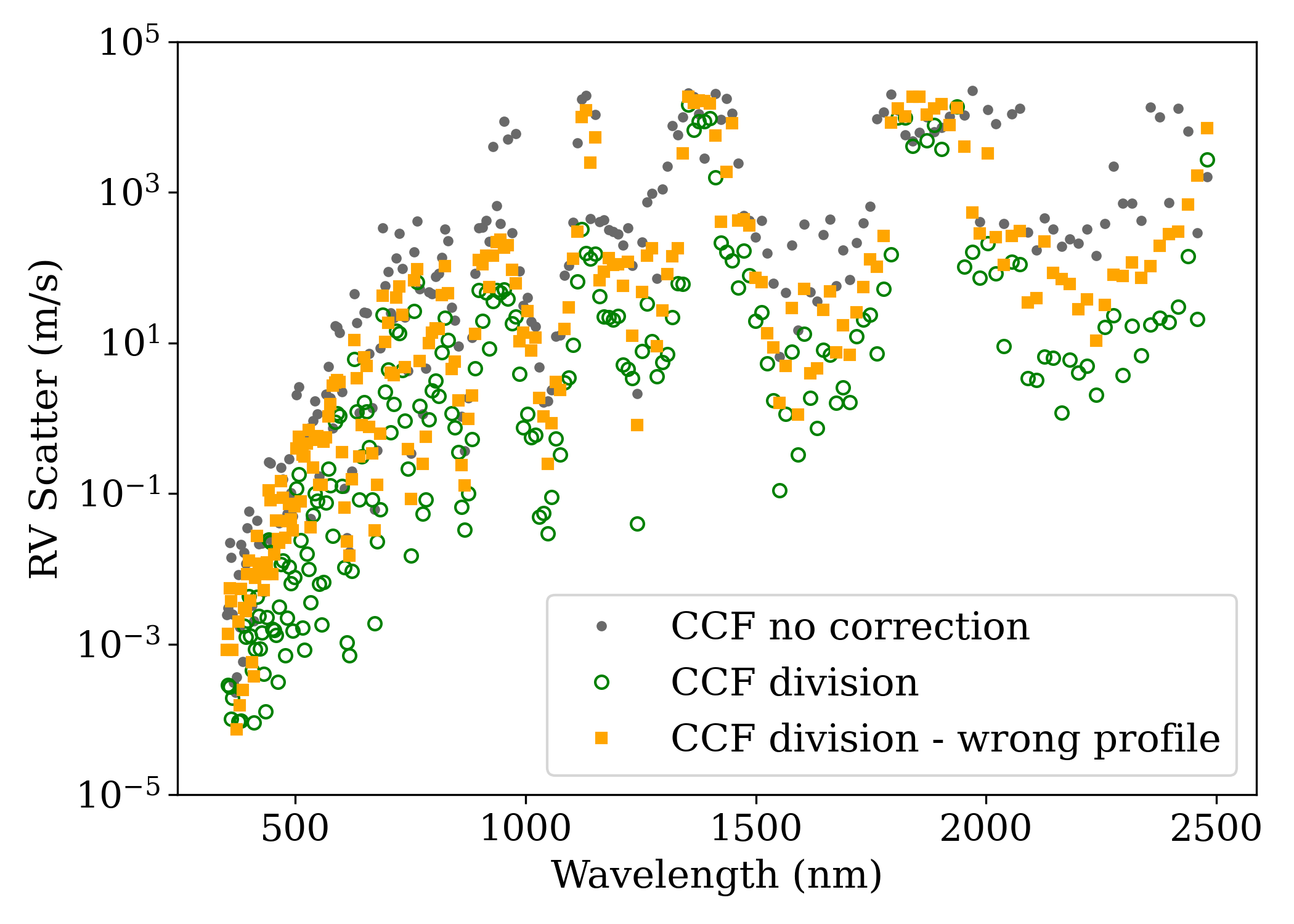}
\includegraphics[scale=0.4]{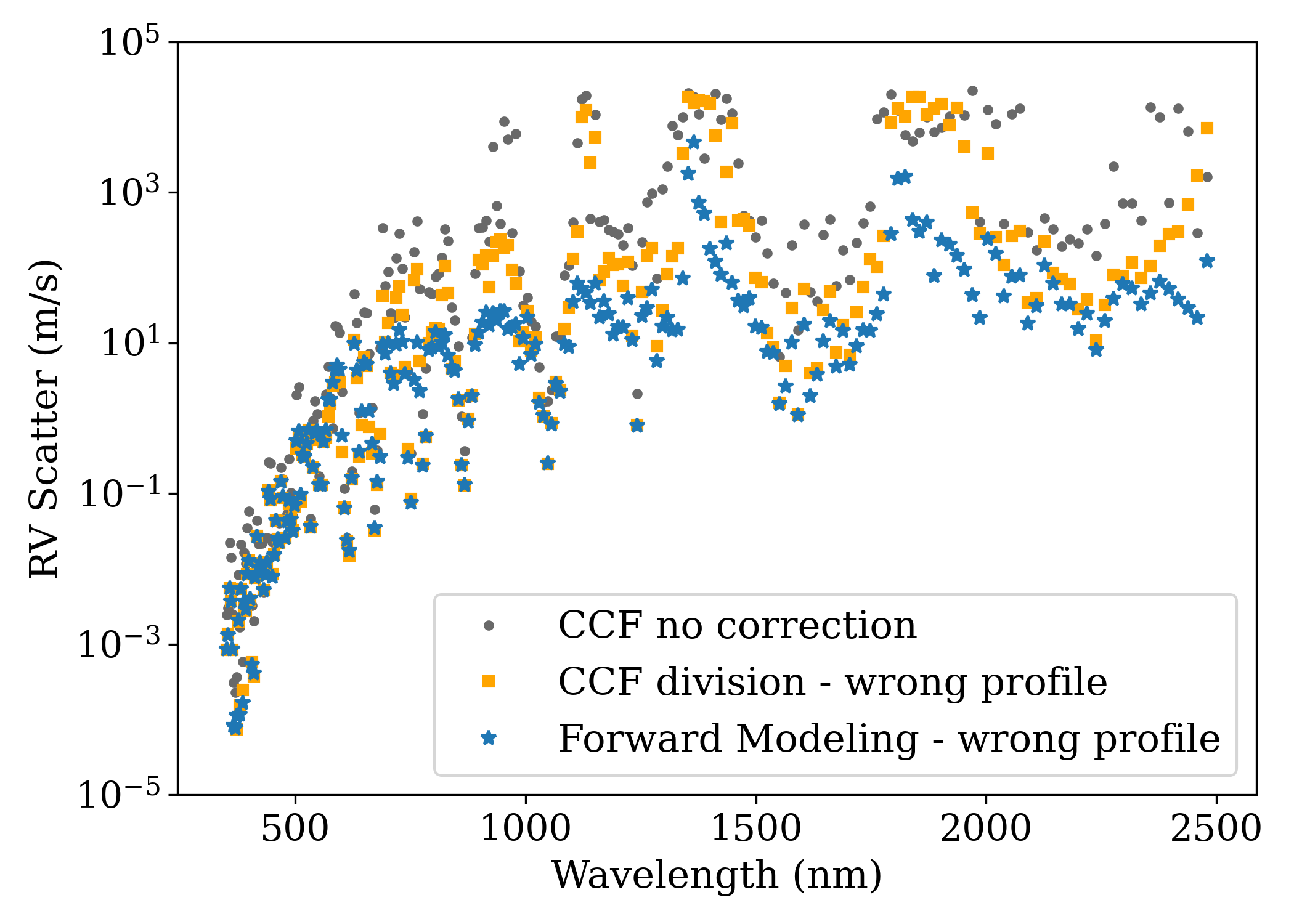} 
\includegraphics[scale=0.4]{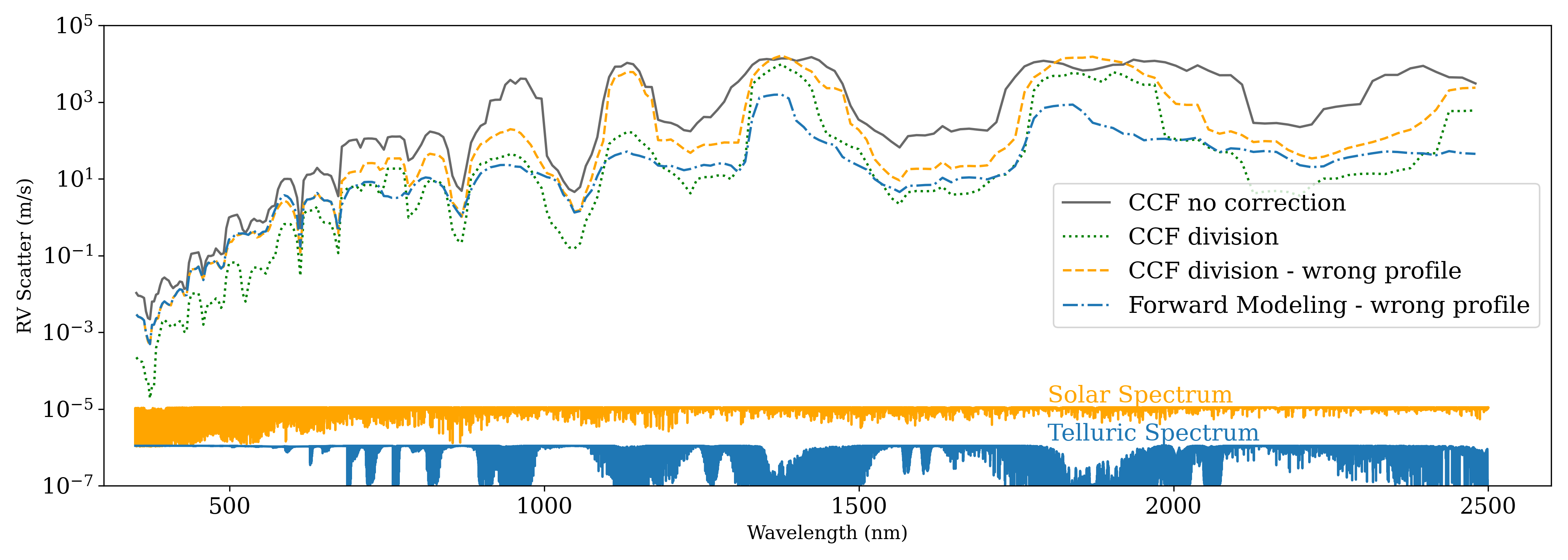}
\caption{RV errors induced by telluric lines for each order in our simulation as a function of wavelengths for various RV extraction methods. Each point in the top two plots is showing the RMS value of the RVs extracted from one spectral order across 365 nights. The top left plot illustrates how using a wrong line profile in the telluric model introduces additional errors. The top right plot illustrates how well the Forward Modeling method out-performs the CCF$+$division method in the NIR. \textbf{RVs extracted using the Forward Modeling method with the correct telluric line profile give the correct answers (0 m/s) across all orders (thus not plotted here).} The bottom plot is a smoothed version of the top plots for clarity (using moving average with a bin size of 5 orders), with the normalized solar (orange) and telluric (blue) spectra plotted at the bottom (scaled arbitrarily, just to show where the lines are).
\label{fig:rmswave}}
\end{figure}

As we divide the stellar spectrum into 230 orders in our simulation and RV extraction, we first take a look at how tellurics affect the RVs in each order. Figure~\ref{fig:chunks} is an illustration of the systematic RV errors when tellurics are being ignored in two different orders -- one in the optical, and one in the NIR. The solar and telluric spectra are plotted on the left to show the contract in spectral Doppler information content between these two orders, and the RVs, as extracted using the CCF method, are plotted as a function of BC on the right. We choose to plot RV vs.~BC instead of day of year because any systematic errors as a result of the line-pulling effect would manifest more directly as a correlation between the RVs and the BCs. This is due to the fact that the BCs dictate the position of the stellar lines on the detector, since usually the bulk of a star's Doppler motion comes from the barycentric motion.

The right panels of Figure~\ref{fig:chunks} show that there are two ways that the tellurics enter the RV error budget, similar to the effects of solar contamination \citep{roy2020}: first, it adds scatter to the measured RVs; and second, it adds a highly coherent systematic signal. This coherent systematic signal has a characteristic width in the wavelength space of a few km/s, which is the typical stellar line width, as a result of the peak pulling effect as the telluric lines ``scan'' back and forth against the stellar lines in the stellar rest frame during the course of one solar year's worth of observations. The amplitude of this systematic signal varies, depending on the competition between relative Doppler content of the stellar lines and the telluric lines. When the stellar lines dominate over the telluric lines, as in the top panels of Figure~\ref{fig:chunks}, the systematic errors are small, and vice versa (as shown in the bottom panels).

We take the standard deviation of the RVs reported for each order over a year as a metric to describe the scatter caused by tellurics at different wavelengths. The three panels in Figure~\ref{fig:rmswave} illustrate the RV scatter as a function of wavelength for various RV extraction methods and telluric mitigation strategies. The top two panels show the RV scatters in each order, with the left panel focusing on the CCF methods and the right panel focusing on comparing the CCF methods with the Forward Modeling methods. Each point plotted is the standard deviation\footnote{We note that using the Median Absolute Deviation (MAD) gives consistent results, and similarly for the RV scatter quoted in the tables and plots in the next subsection as well.} of the RVs reported by a single order across a year (i.e., the scatter in the right panels of Figure~\ref{fig:chunks}). Overall, the amplitudes of the RV scatters trace the telluric bands -- in wavelength regions where the telluric absorption is heavy (e.g., the places between the NIR photometric bands), the RV scatter is significantly larger than the others. The optical region experiences less RV scatter than the NIR, as expected.



Dividing out the telluric absorption lines clearly improves the RV precision across the entire spectrum, but the CCF division method does not remove the effects of tellurics completely, even though it was a ``perfect'' division. Here, the telluric lines were divided out ``perfectly'' in the sense that what was divided out from each observed spectrum was the exact same input telluric spectrum used when simulating the observed spectrum (convolved with the known spectral PSF to match the line widths). This is, in fact, not a perfect division because of convolution, as discussed in the very beginning of this section, Section~\ref{sec:results}. As a result of this residual ``convolution-division'' error, the RV scatter was not reduced to zero even though a ``perfect'' telluric model was used. 


The top right panel of Figure~\ref{fig:rmswave} compares the RV scatters using the Forward Modeling method with the CCF method, both using wrong line profiles. The RV scatters of the no correction method are also plotted as a fiducial baseline for comparison. Note that we did not plot the results for the Forward Modeling method with no correction, because it produces the same results as the CCF with no treatment method, just as expected. The results for the Forward Modeling with the perfect telluric model are also not plotted, because this method produces essentially 0~m/s (within machine precision) for all orders across all simulated spectra (i.e., it completely eliminates the effects of telluric contamination). It outperforms the CCF with perfect division because Forward Modeling is not subject to the convolution-division error. For this same reason, the Forward Modeling with wrong profiles method outperforms CCF with wrong profiles in a typical order with moderate to deep telluric lines.

The bottom panel of Figure~\ref{fig:rmswave} shows an overall comparison for all methods, with the RV scatter values now smoothed over five orders and plotted as lines for clarity. In particular, CCF division (with perfect lines) outperforms Forward Modeling with wrong profiles in regions where telluric lines are shallow, especially in the optical region. This is because when the telluric lines are shallow, the convolution-division error is small compared with the residual error caused by the wrong line profile. As shown in Figure~\ref{fig:conv}, the RMS of the convolution-division error is much smaller than 1\% for shallow lines (and scales roughly with line depths), while the line profile mismatch induces a typical RMS of about 1\% (Figure~\ref{fig:lineprofile}). For a similar reason, the CCF division and the Forward Modeling methods essentially perform equally well in the optical region bluer than 600~nm when using the wrong line profile. Another reason why the CCF division (with perfect lines) performs very well than the other methods in the optical region blue-ward of 700~nm is that the continuum absorption caused by O$_3$ was divided out completely, as we used the perfect input telluric model. In contrast, the ozone absorption was ignored in the CCF division or Forward Modeling methods with the wrong line profile, as it would be unrealistic to constrain the absorption of ozone, which is equivalent to a very small change in the continuum level in the observed spectrum. We remind the readers again that the best performing method, which cannot be plotted in Figure~\ref{fig:rmswave}, is the Forward Modeling method when the perfect telluric model is employed, because all RV errors induced by the tellurics are eliminated.

\subsection{What Correlates with Telluric Induced RV Errors}

As shown in Figure~\ref{fig:rmswave}, the impact of telluric contamination on the RVs varies across wavelengths because of how telluric lines distribute differently in different bands, across over seven orders of magnitude from sub-cm/s to over km/s. In reality, the RV precision from any single order is often not high enough due to limited SNR, so RVs across multiple orders are combined to report a final RV for each night. In this section, we examine the effects of telluric contamination on the combined, final RVs. How the RVs are combined is described in Section~\ref{sec:method}, and briefly, it is a weighted average with the weights estimated using the RV precision limit set by the photon-limited Doppler content and the RV scatter caused by tellurics.

\begin{figure}[]
\includegraphics[scale=0.38]{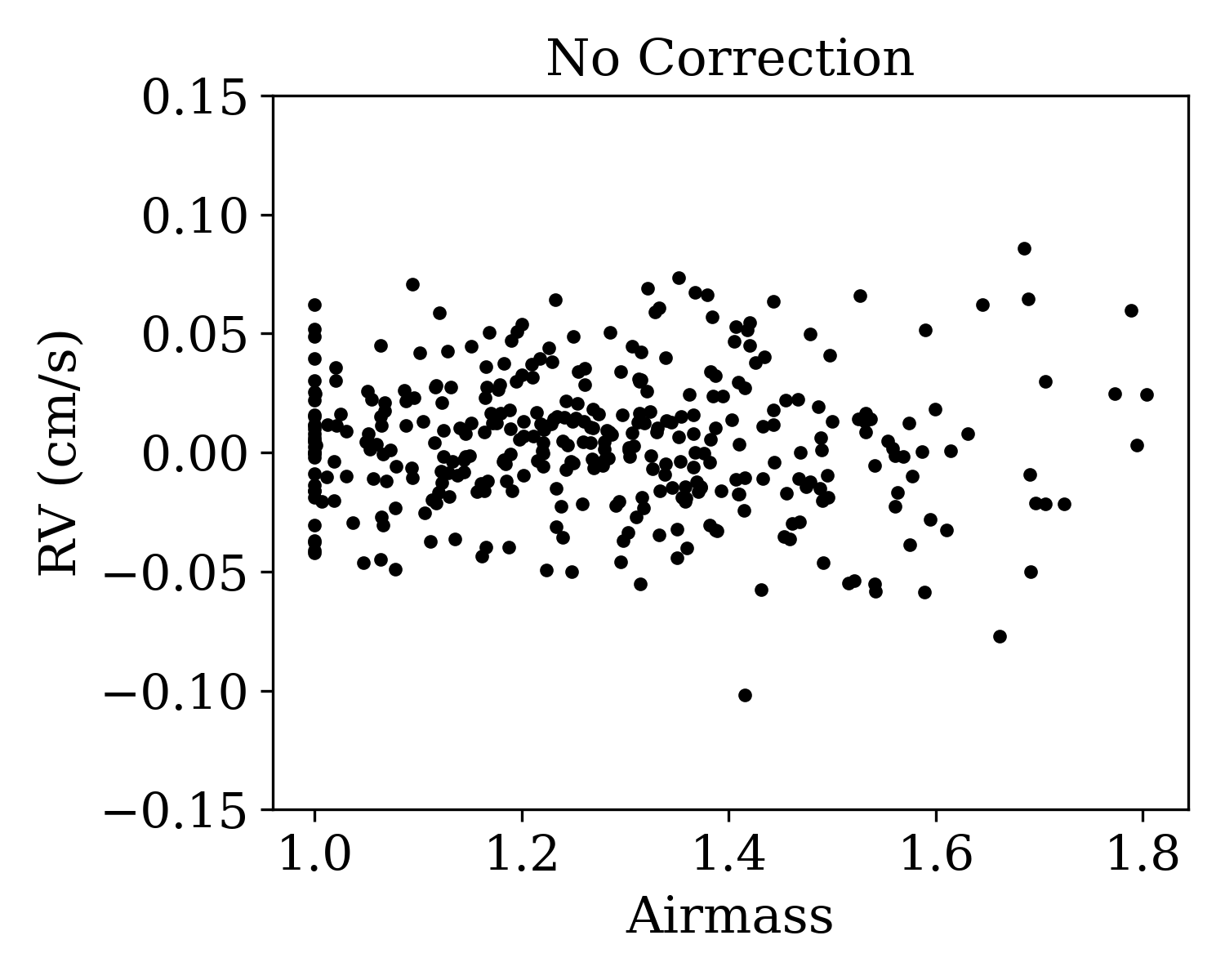}
\includegraphics[scale=0.38]{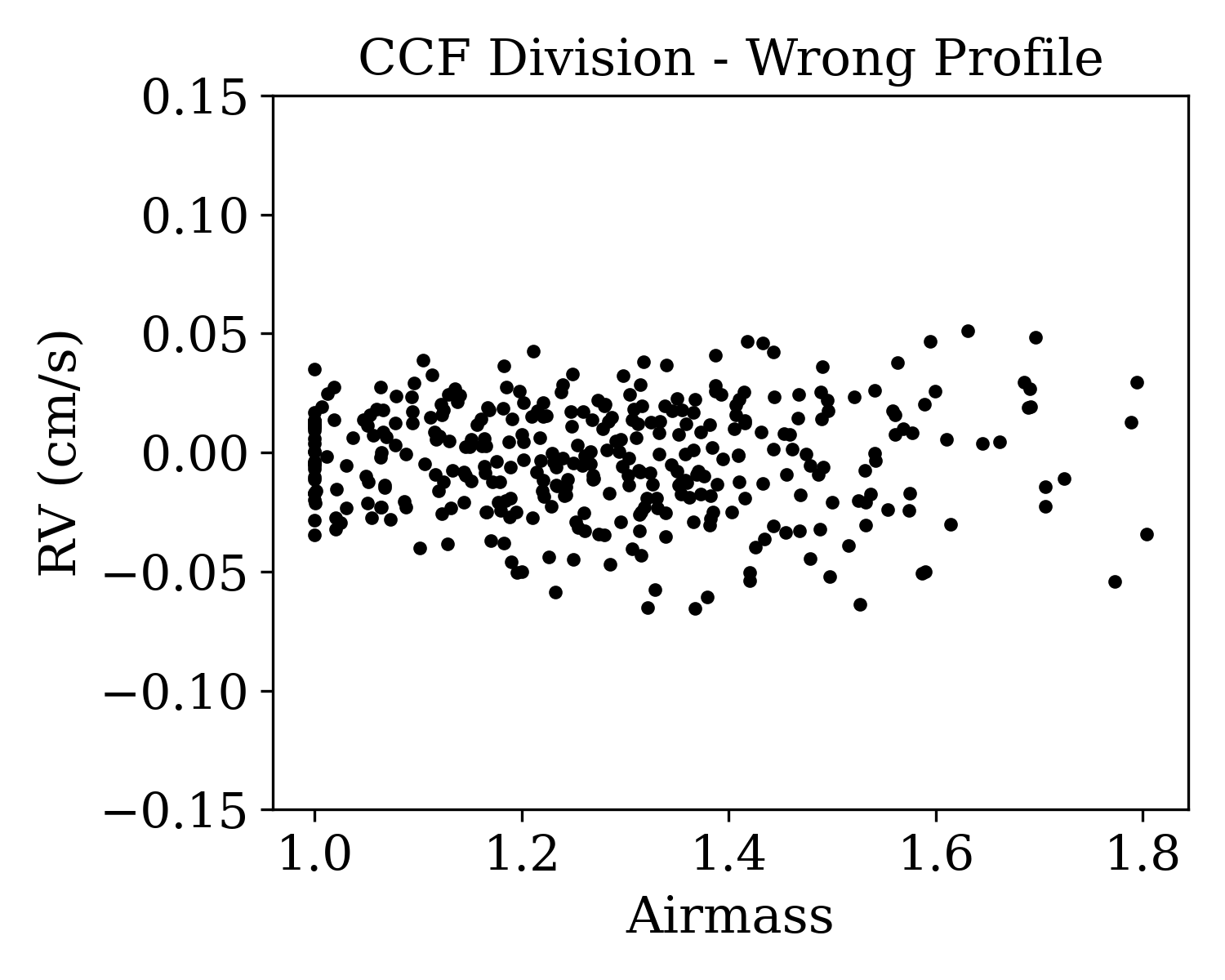}
\includegraphics[scale=0.38]{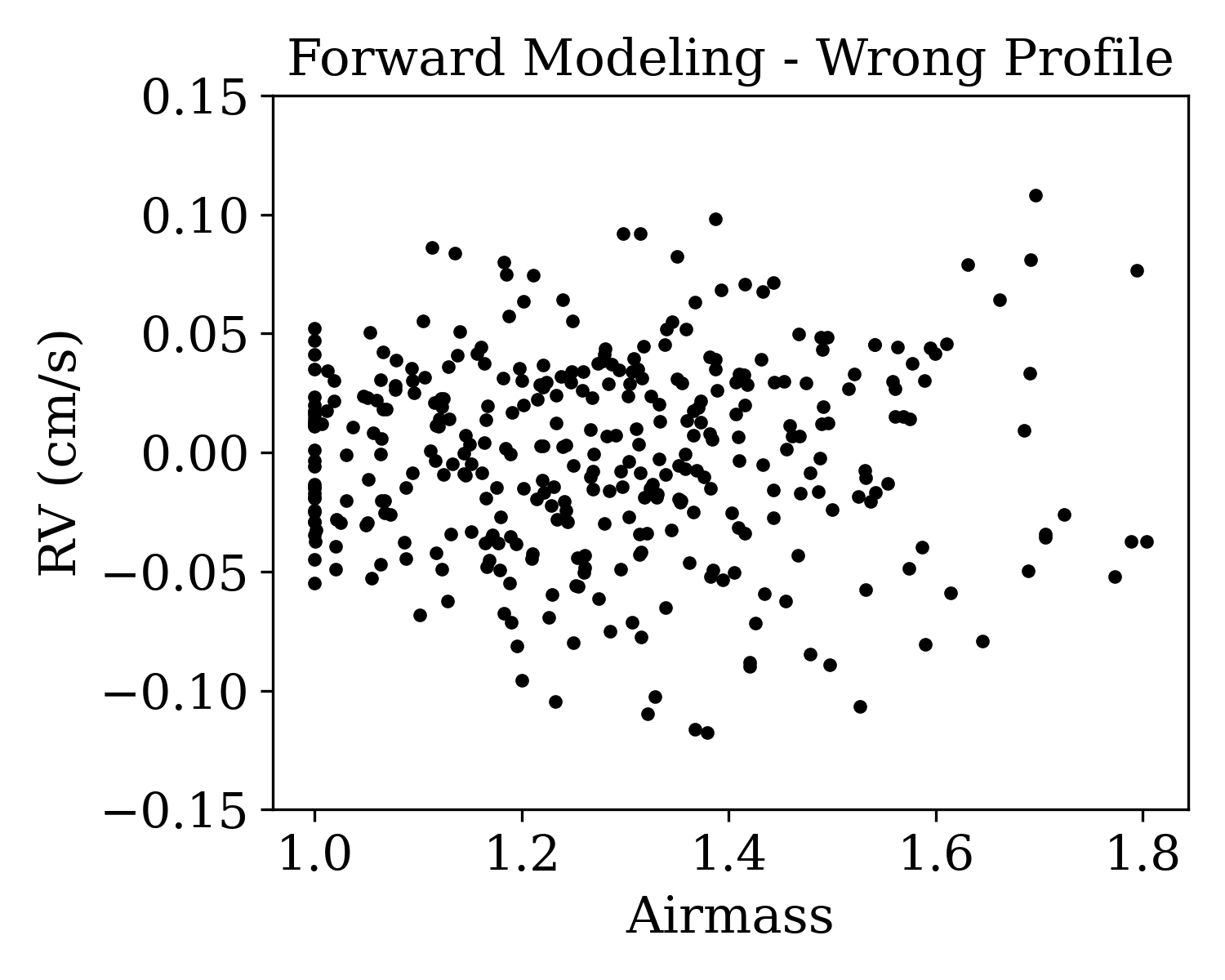} \\
\includegraphics[scale=0.38]{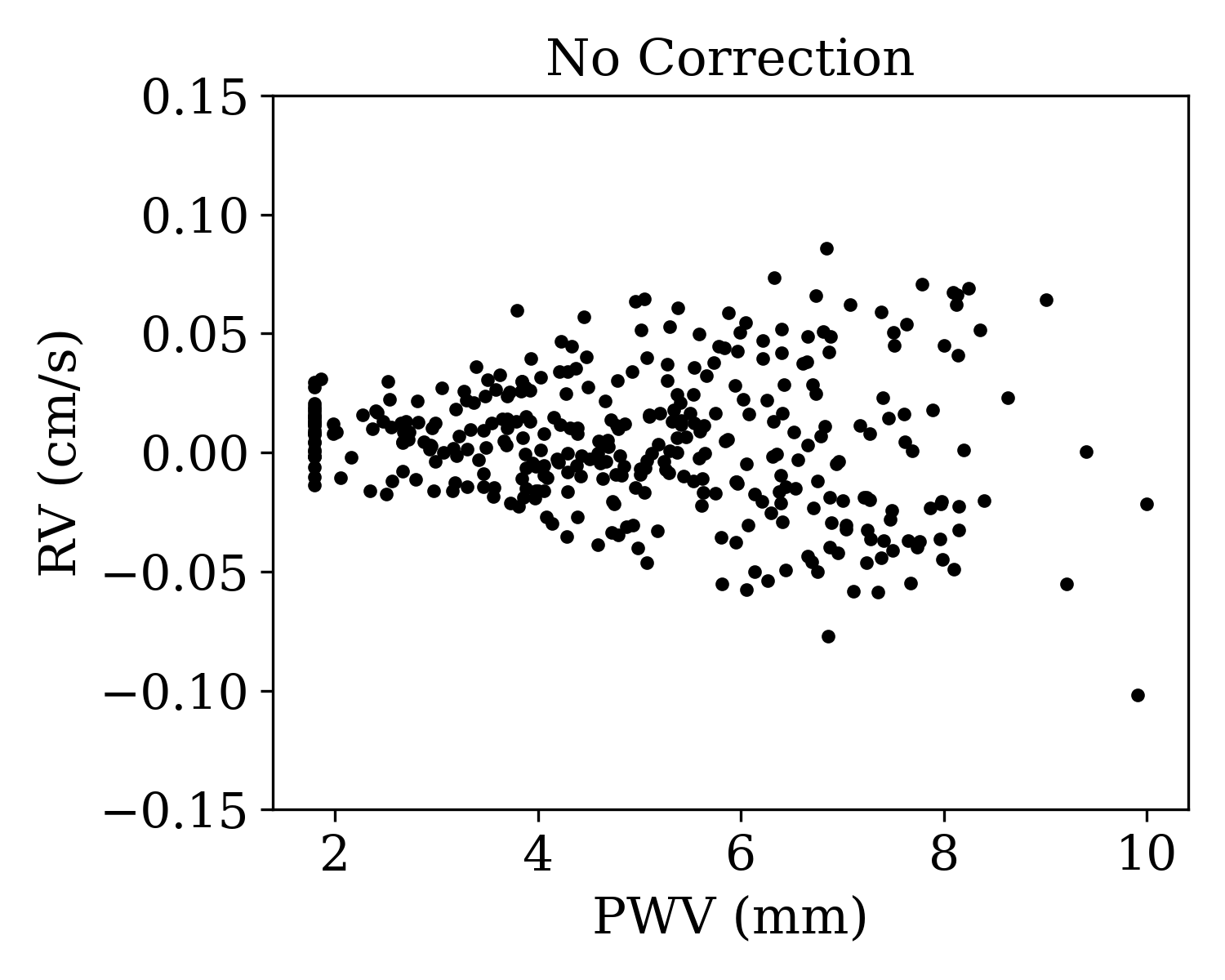} 
\includegraphics[scale=0.38]{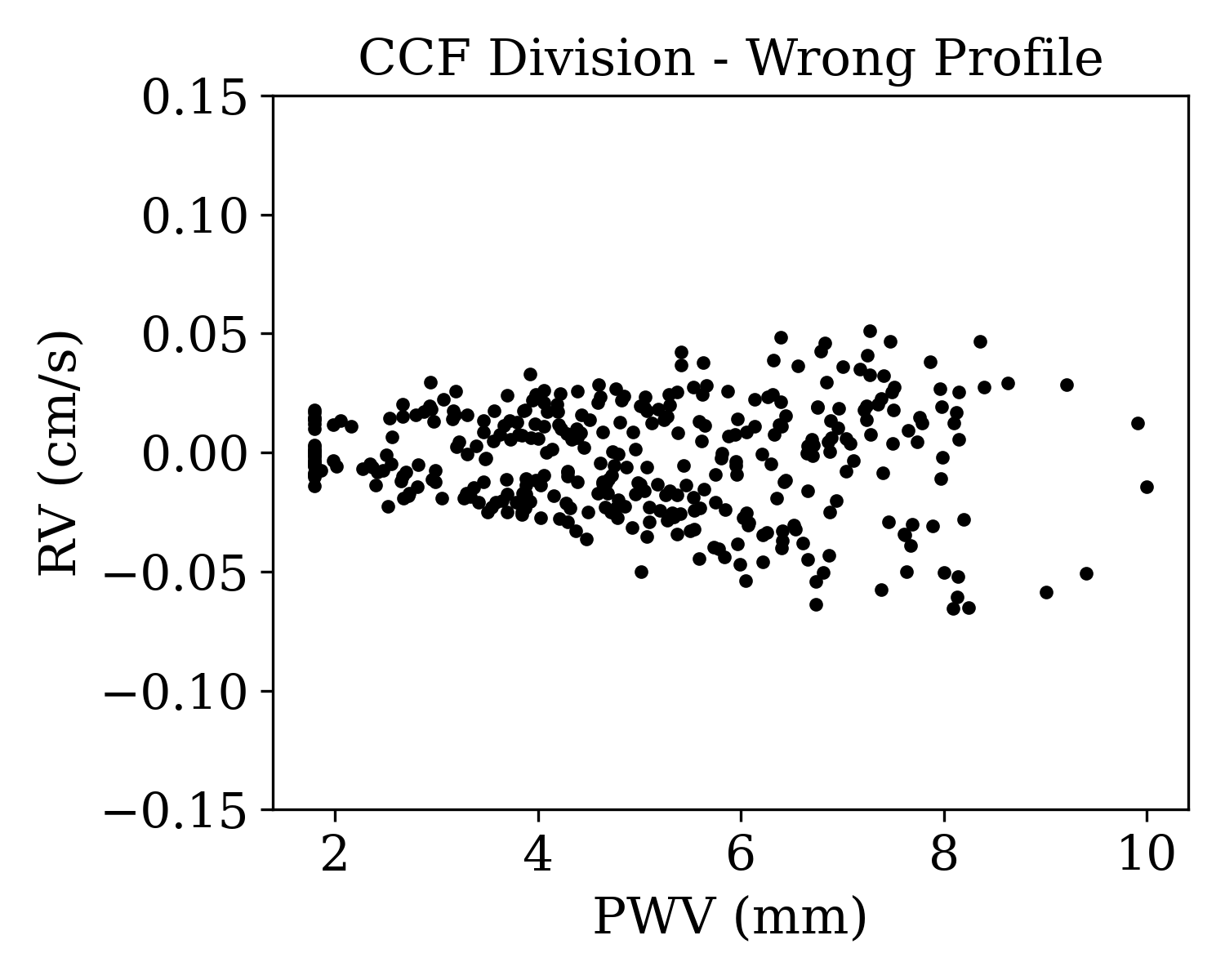} 
\includegraphics[scale=0.38]{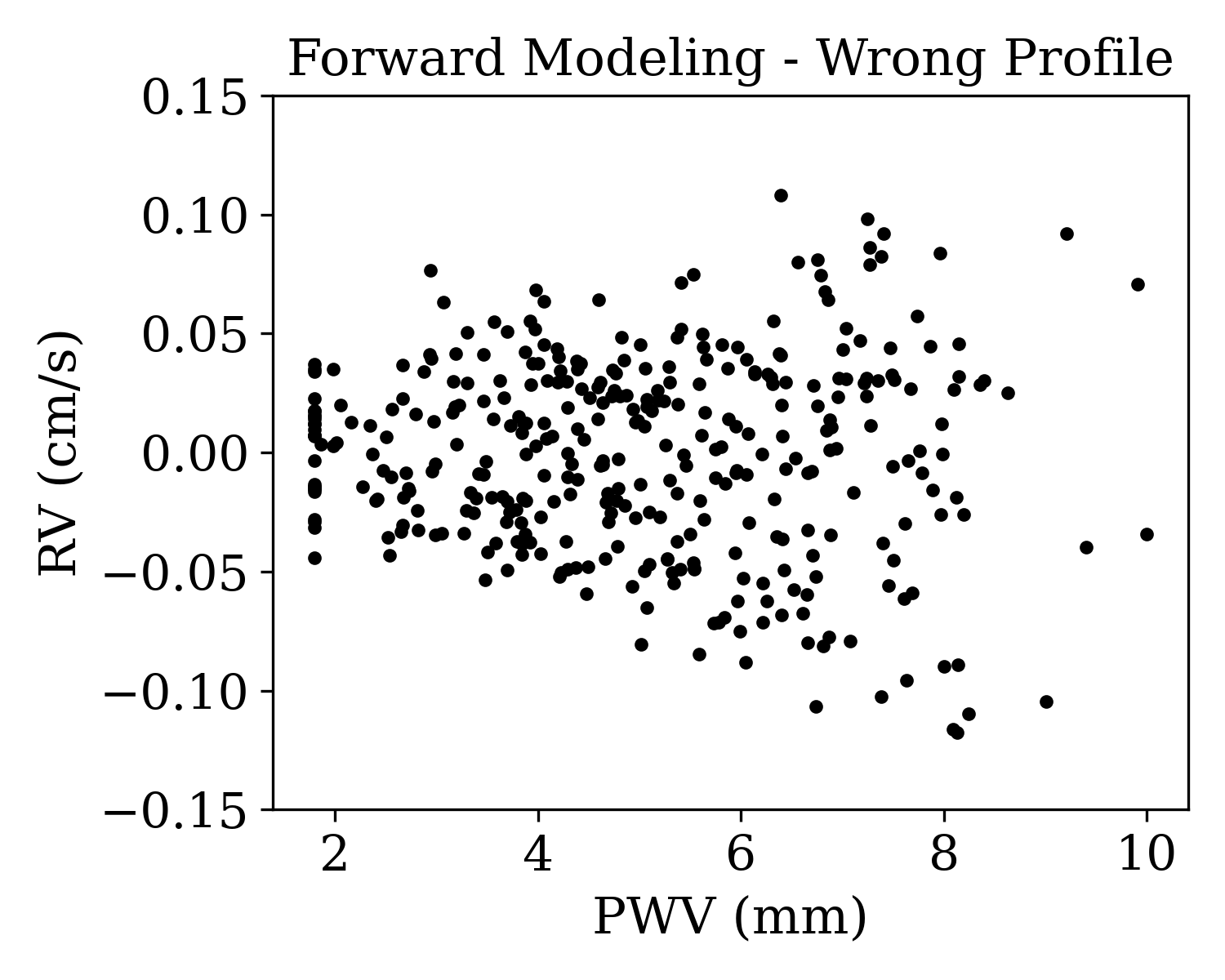} \\
\includegraphics[scale=0.38]{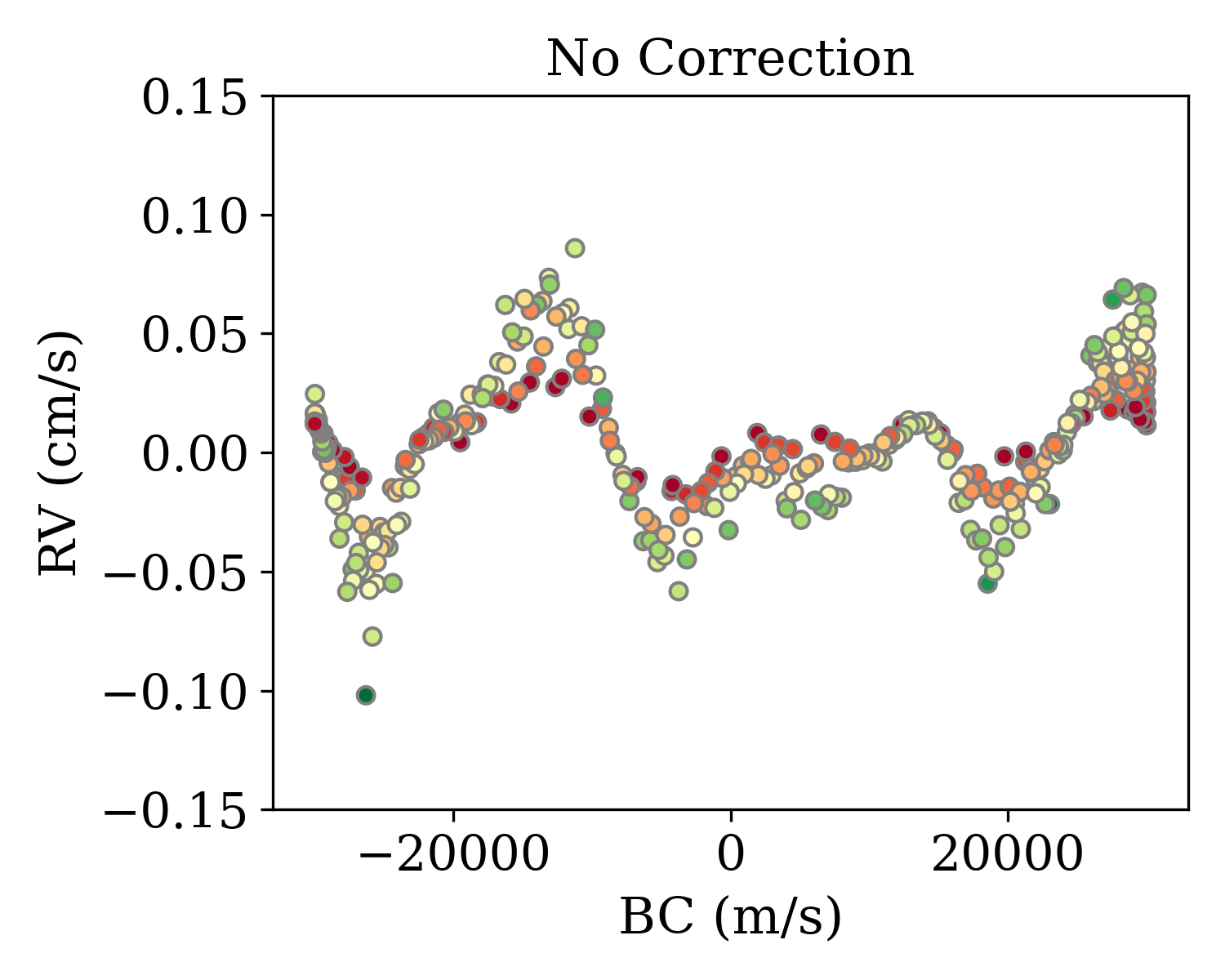} 
\includegraphics[scale=0.38]{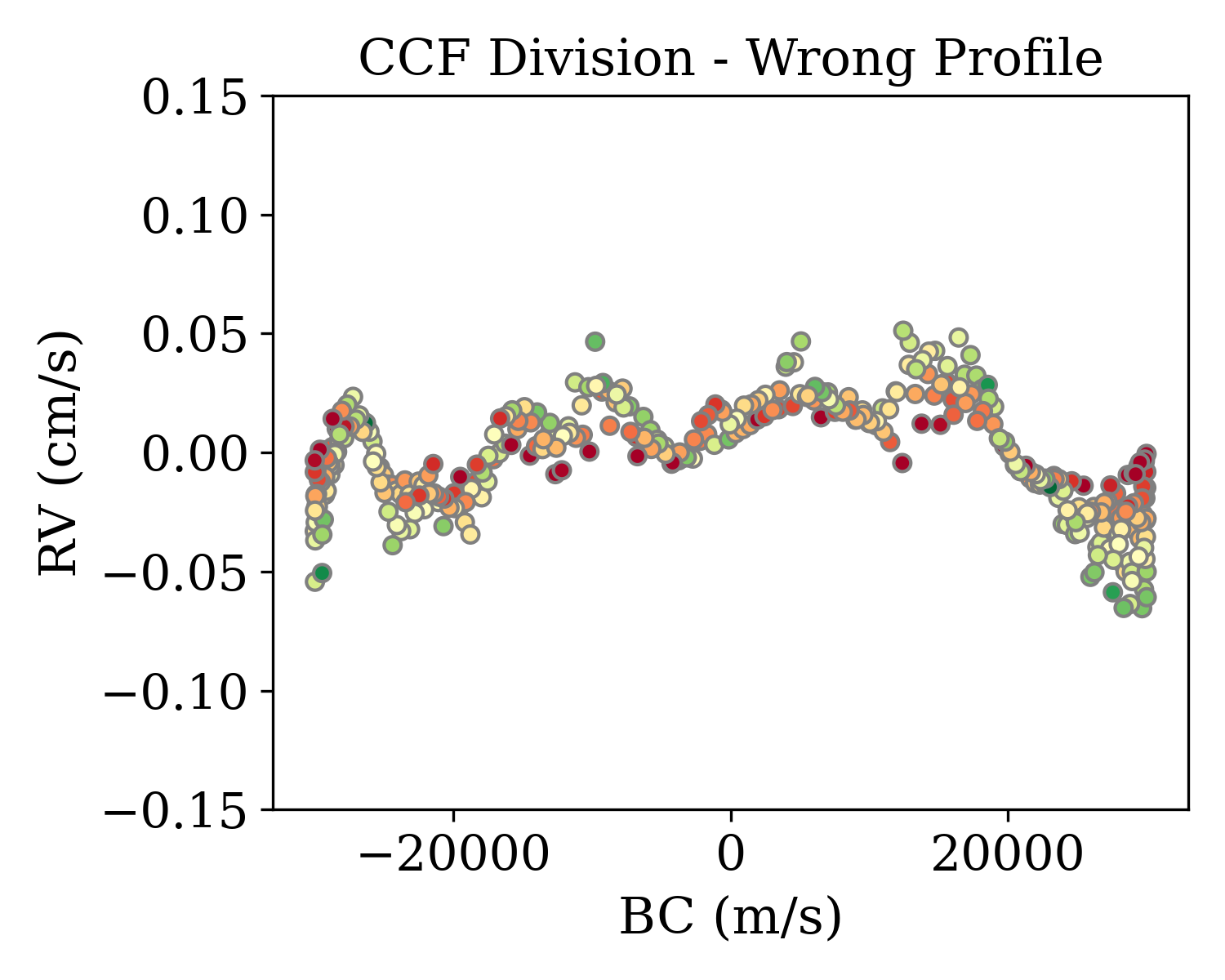} 
\includegraphics[scale=0.38]{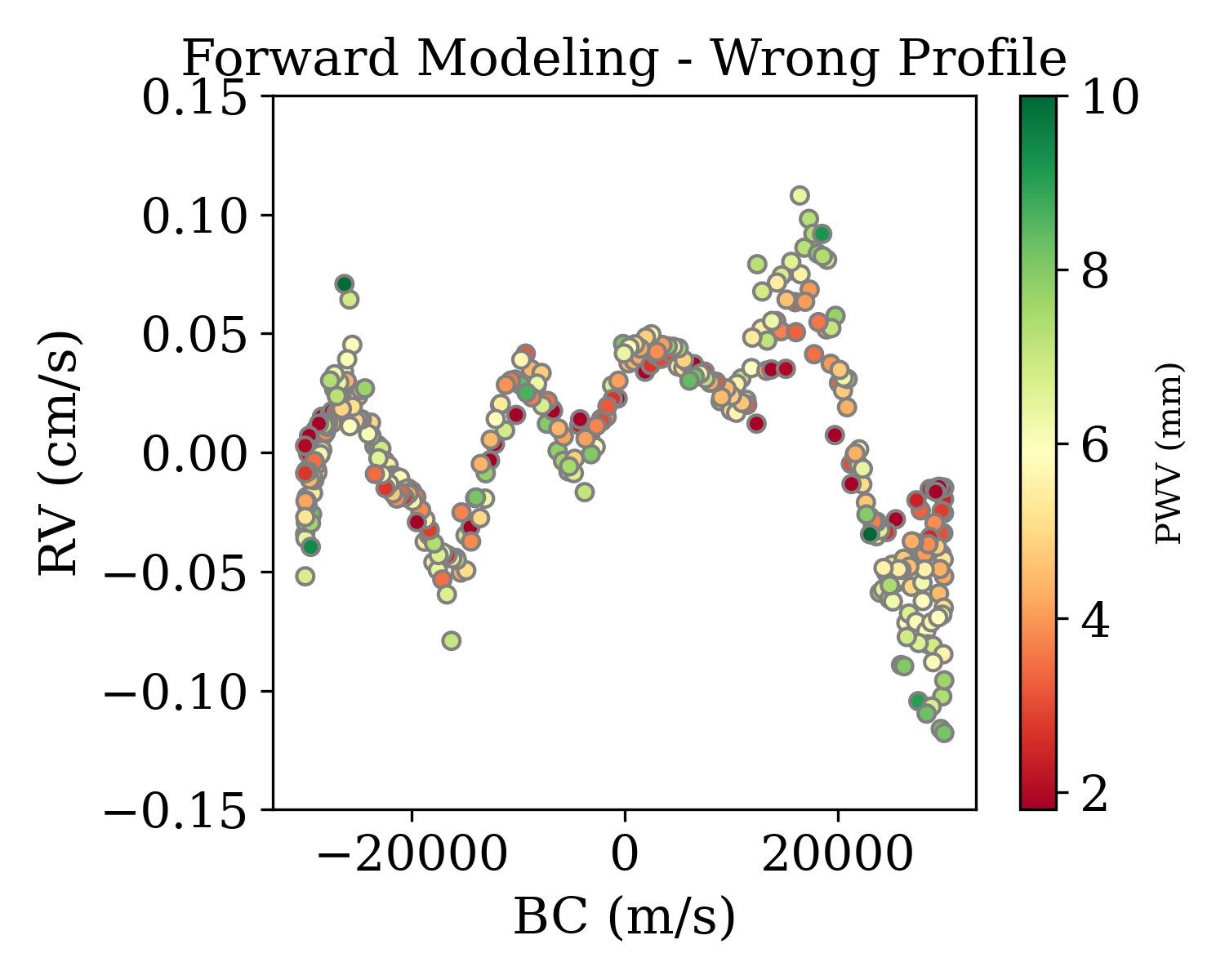} 
\caption{Extracted RVs for each night as a function of airmass, PWV, and BC for three different methods of RV extraction (left: no correction; middle: CCF$+$division with a wrong telluric line profile; right: Forward Modeling with a wrong telluric line profile). Each RV data point is a weighted average of the RVs reported by all orders in the entire wavelength range used in our simulation (350~nm -- 2.5 $\mu$m). The bottom panels are color-coded by the PWV values in mm. The weights are a combination of photon noise (assuming SNR = 100 per pixel) and RV RMS induced by tellurics, as described in Section~\ref{sec:method} and same as used for data presented in Table~\ref{tab:rvinst}. Water absorption clearly plays an important role in the final RVs, as shown by the strong correlation between RVs and PWVs. Similar to the right plots in Figure~\ref{fig:chunks}, the extracted RVs have strong temporal structures caused by changes in the separation between telluric lines and stellar lines in wavelength space due to the Earth's barycentric motion. 
\label{fig:pabcalc}}
\end{figure}

We first look at how RVs are correlated with the airmass and the PWV, which, to the first order, correlate with telluric line depths. The first two rows of plots in Figure~\ref{fig:pabcalc} show the correlation between the final RV (a weighted sum across our entire spectral range) versus the airmass or PWV for each night. The plots are for three different RV extraction methods: no correction (again as the fiducial baseline of comparison), CCF with the wrong profiles, and Forward Modeling with wrong profiles. The RVs from the CCF division with the perfect tellurics show similar patterns, and the Forward Modeling method with perfect lines reports RVs equal to 0 across all wavelengths and nights. Therefore, we choose the two methods with wrong profiles as a representative set of results. 

The top plots of Figure~\ref{fig:pabcalc} show that the amount of scatter in the RVs is only very weakly correlated with the airmass. This is because the final RVs are dominated by the RVs reported in the optical region (350--600~nm), especially the bluer part, with ample solar lines and few telluric lines in comparison. The telluric lines in the blue/green optical region are basically all water lines, whose depths are mostly governed by the PWV. PWV variations across nights can be easily more than factor of two, much more than the change in airmass. This is also reflected in the middle plots of Figure~\ref{fig:pabcalc}, where the RV scatter is clearly correlated strongly with the PWV.

In the bottom plots of Figure~\ref{fig:pabcalc}, we show the final RVs versus the barycentric velocity, BC, with the PWV color coded. The telluric contamination induces a small coherent systematic error with an amplitude of $\sim$cm/s in the RVs from all three methods. BC stands out as the most clear and strongest correlated with the RV error, with PWV as the second one, explaining the amplitude of the RV error at any given BC. Airmass does not seem to have a strong correlation with the RVs even in the regions where there is no water lines (i.e., if we plot the same set of plots with RVs derived using spectra from a water-free region instead of our entire wavelength range, the top plots would still look similar).

There are two reasons why the amplitude of this systematic error did not decrease significantly with our treatment of telluric lines, and it might look surprising that the amplitude of the RV systematics in the Forward Modeling method (lower right panel) is larger than the no correction or CCF division method. This is due to our adopted weighting scheme. First, the final RVs are dominated by the best-behaving orders in the blue/green part of the optical region, where CCF or Forward Modeling with wrong profiles perform similarly to the CCF no correction method, as shown in the bottom panel of Figure~\ref{fig:rmswave}. Second, CCF or Forward Modeling with wrong profiles did not correct for the continuum absorption in the optical region caused by the ozone, which is the dominant source of systematics when water lines are absent in the blue. This also means that our results and conclusions would basically remain unchanged if we use a different set of randomly generated PWV and airmass values. 

Why do the systematic patterns from different orders not cancel out after combining the RVs? As illustrated in Figure~\ref{fig:chunks}, different orders exhibit different systematic errors with varying patterns and amplitudes. When combining the RVs from different orders to report a final RV on any given night, unfortunately, these scatters and systematics do not cancel. The characteristic width of these systematic signals are similar, though not the same, across all orders, since it is tied to the line widths. Therefore, when adding all these systematic errors together, they persist as a coherent systematic error as well. This is analogous to adding a finite series of quasi-periodic curves with similar periods but varying amplitudes and phases, and for the weighted average of them to have a coherent structure as well (with a reduced amplitude).

The high correlation between RV, BC, and PWV shown in the lower left panel of Figure~\ref{fig:pabcalc} is intriguing: this means that if the stellar spectrum and the telluric absorption lines in a series of observations are precisely known, then the RV bias induced by telluric could in principle be predicted via simulations like this work. However, this could be hard to implement in reality as it is hard to know the stellar and telluric spectra very well, but it nonetheless presents a possible alternative path forward in mitigating tellurics.

\subsection{Tellurics' Contribution To The RV Error Budget in Different Photometric Bands}\label{subsec:errorbands}

Next, under our assumptions when combining the RVs, we examine the total amount of scatter in the RVs induced by tellurics over a wavelength range. We first see how much the telluric contamination would contribute to the RV error budget for several new generation high precision RV instruments, as listed in Table~\ref{tab:rvinst}. We caution the readers that there are several important assumptions here, which would make such estimates not accurate for these instruments but just serving as a comparison between them in stead. First, we did not consider the various instrumental resolutions, which typically differ from the resolution of our simulated spectra. Second, we assume an SNR = 100 per pixel in our simulated spectra when combining the RVs, while real observations would have a variety of SNRs (especially that the resolution and the pixel sampling factors of different instruments also differ). The RV scatters listed are simply the standard deviation in the combined RVs over the wavelength range of each instrument.

Comparing the first column, ``No Correction'', with the last column listing the photon-limited precision in Table~\ref{tab:rvinst}, it is evident that for the instruments operating in the broad optical region, tellurics do not contribute significantly to the error budget, assuming proper weighting is applied when combining the RVs. In the NIR, however, tellurics affect the RV precision significantly. Even after dividing out or modeling the tellurics, the RV precision still could not reach the photon-limited precision, with a scatter of about 1~m/s. Overall, the three mitigation methods perform on a similar level, with CCF division with perfect tellurics outperforms the other two (again recall that Forward Modeling with perfect tellurics eliminates the telluric-induced errors completely). This is consistent with the results shown in Figure~\ref{fig:rmswave}, where CCF division with perfect tellurics performs the best in most orders, especially the ones with relatively shallower telluric lines.

\begin{deluxetable}{cccccc}
	\tablecaption{Summary of the RV scatters for different instruments using different RV extraction methods. EFE stands for EarthFinder Equivalent on the ground. The RVs from different orders were combined using weighted average taking into account both the scatter caused by the tellurics and the photon-limited RV precision assuming an SNR of 100 per pixel. \label{tab:rvinst}}
    \tablehead{
    \colhead{Ground-based} & \colhead{No Correction} & \colhead{Division} & \colhead{Division} & \colhead{Modeling} & \colhead{Photon-limited} \\
    \colhead{Instrument} & \colhead{(m/s)} & \colhead{(m/s)} & \colhead{K-Profile (m/s)} & \colhead{K-Profile (m/s)} & \colhead{Precision (m/s)}
    }
    \startdata
      EFE Visible Arm & & & \\ (380-900 nm) & 0.035 &   0.021 &   0.021 &   0.032 &   0.059 \\ \hline
      EFE NIR Arm &&&&\\
      (900-2500 nm) & 2.432 &   0.761 &   1.196 &   1.220 &   0.143\\ \hline
      ESPRESSO &&&&\\(380-788 nm) & 0.034 &   0.020 &   0.018 &   0.023 &   0.062 \\ \hline
      EXPRES &&&&\\(380-680 nm) & 0.031 &   0.013 &   0.021 &   0.024 &   0.061 \\ \hline
      NEID &&&& \\(380-930 nm) & 0.034 &   0.019 &   0.021 &   0.033 &   0.059 \\ \hline
      CARMENES &&&& \\Visible Arm &&&& \\(520-960 nm) & 0.169 &   0.098 &   0.111 &   0.179 &   0.107 \\ \hline
      CARMENES &&&& \\NIR Arm &&&& \\(960-1710 nm) & 2.359 &   0.659 &   1.161 &   1.144 &   0.229 \\ \hline
     \enddata
\end{deluxetable}

If we take a more careful look and see how the tellurics affect the RV precision in narrower bands, we would see that it is not all bad in the NIR. Table~\ref{tab:rvband} lists the RV scatter caused by tellurics in several commonly used photometric bands. I, Z and Y bands in the short NIR could actually reach very close to the photon-limited precision after treating the tellurics. Interestingly, in these three bands, modeling with wrong profile is worse than division with wrong profile in certain bands, which seems at odds with the results for different orders in Figure~\ref{fig:rmswave}. This is because of weighting. 

In the I, Z and Y bands, the division performs as well as the modeling in regions almost free of tellurics. In regions with tellurics, division is significantly worse, so these orders receive very low weights in Division. However, the same orders received higher weights as they were not as worse and sneaked into the final RVs of Modeling, making its RV RMS worse. For example, in the Z band, there are several orders that are reporting $\sim$m/s precision (840--900~nm) for both division and modeling, and the rest of the orders are typically ~10 m/s for modeling, but $\sim$50$+$ m/s for division. If we put more equal weights between the orders or weight the orders only by their RV scatter from tellurics, modeling would report better results than division, as expected by looking at Figure~\ref{fig:rmswave}. However, such a weighting scheme is rarely adopted in real practice: equal weighting is certainly far from optimized, and weighting only by the telluric-induced errors is unrealistic, because it is hard to distinguish from the total RV scatter how much is the true photon-limited error versus the residual impact of tellurics in real observations. Still, the fact that Forward Modeling does not out-perform the CCF method, as we would have expected based on Figure~\ref{fig:rmswave}, suggests that our weighting scheme, a commonly adopted in real practice, can be improved.

\begin{deluxetable}{cccccc}
	\tablecaption{Summary of RV RMS precision lower limits for several common photometric bands. The RVs were combined in the same fashion as in Table~\ref{tab:rvinst}.\label{tab:rvband}}
    \tablehead{
    \colhead{Photometric} & \colhead{No Correction} & \colhead{Division} & \colhead{Division} & \colhead{Modeling} & \colhead{Photon-limited} \\
    \colhead{Band} & \colhead{(m/s)} & \colhead{(m/s)} & \colhead{K-Profile (m/s)} & \colhead{K-Profile (m/s)} & \colhead{Precision (m/s)}
    }
    \startdata
      B (398--492 nm) & 0.022 &   0.001 &   0.008 &   0.008 &   0.094\\ \hline
      V (507--595 nm) & 0.167 &   0.036 &   0.072 &   0.072 &   0.219\\ \hline
      R (589--727 nm) & 0.446 &   0.283 &   0.358 &   0.454 &   0.285\\ \hline
      I (732--881 nm) & 0.968 &   0.427 &   0.645 &   0.746 &   0.396\\ \hline
      Z (0.8--1.0 $\mu$m) & 1.283 &   0.790 &   0.541 &   1.210 &   0.416\\ \hline
      Y (0.96--1.1 $\mu$m) & 2.414 &   0.778 &   1.272 &   1.217 &   0.993\\ \hline
      J (1.1-1.3 $\mu$m) & 5.816 &   2.821 &   5.193 &   3.473 &   1.027\\ \hline
      H (1.5--1.8 $\mu$m) & 5.789 &   1.169 &   1.597 &   1.586 &   0.610\\ \hline
      K (2.0--2.4 $\mu$m) & 58.358 &   2.187 &   7.686 &   6.778 &   1.093\\ \hline
     \enddata
\end{deluxetable}

\begin{figure}[ht]
\centering
\includegraphics[scale=1.0]{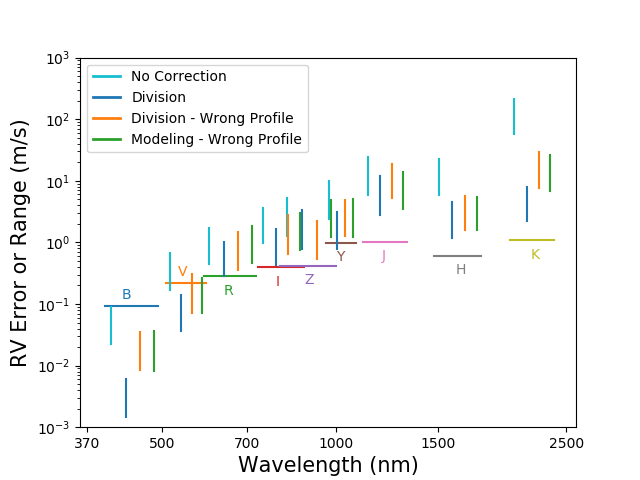}
\caption{Telluric induced RV errors for each photometric band. The horizontal lines are marking the photon-limited precision in each band (for SNR=100 per pixel) and the wavelength range of each band. Each vertical line marks the standard deviation in the RVs (lower end of line; same as in Table~\ref{tab:rvband}) and the range of the RV systematics (95\% percentile range; upper end of line) from each of the four simulation methods as listed in Table~\ref{tab:rvband} (plotted in groups of four vertical lines within the wavelength range of each band).
\label{fig:banderror}}
\end{figure}

Both Table~\ref{tab:rvinst} and \ref{tab:rvband} list the standard deviation of the combined RVs. If we use the Median Absolute Deviation instead of the standard deviation, the scatter would decrease considerably, up to a factor of two. This is not surprising since the RV scatter is not Gaussian, but a systematic pattern as shown in the bottom panels of Figure~\ref{fig:pabcalc}. Therefore, different summary statistics would report a different value for the RV scatter. To capture this complication and to provide a straight-forward illustration, we summarize the RV variations across different photometric bands in Figure~\ref{fig:banderror}, where both the range of the RV systematics (95\% percentile) and the standard deviation of RVs in any band are plotted as the two ends of each vertical line.

\subsection{How Telluric Induced RV Errors Manifest in Time Series}\label{sec:timeseries}

\begin{figure}[!th]
\centering
\includegraphics[scale=0.55]{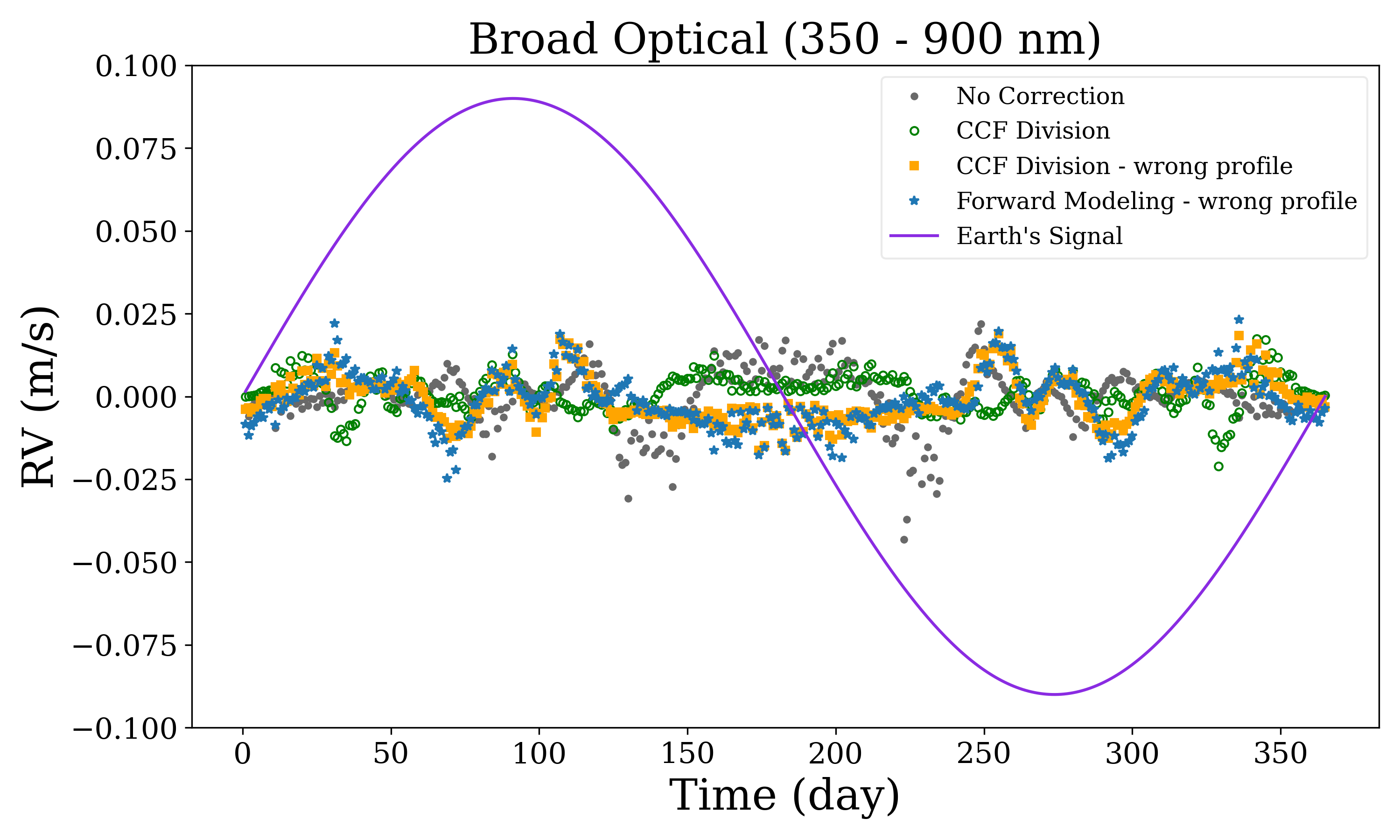}
\includegraphics[scale=0.55]{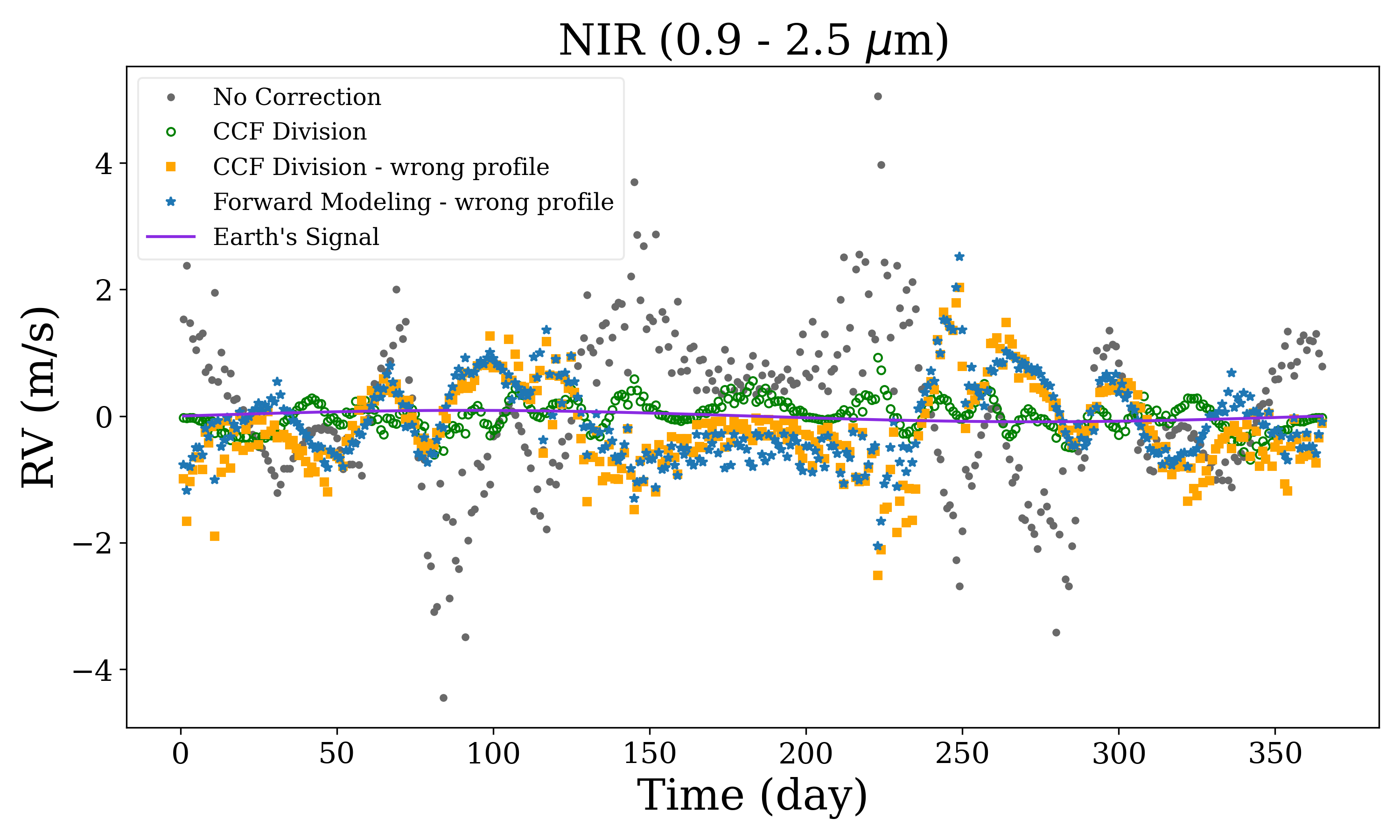}
\caption{RV signals vs.\ time, as measured by the visible arm (upper panel) and NIR arm (lower panel) of an 
EarthFinder equivalent (EFE) spectrograph from the ground, extracted with four different telluric mitigation methods, assuming SNR=400 per pixel for R=120,000. The RV signal of an Earth analog is plotted as the purple line (semi-amplitude = 9 cm/s). See Section~\ref{sec:timeseries} for more detail.
\label{fig:rvtime}}
\end{figure}

\begin{figure}[!th]
\centering
\includegraphics[scale=0.55]{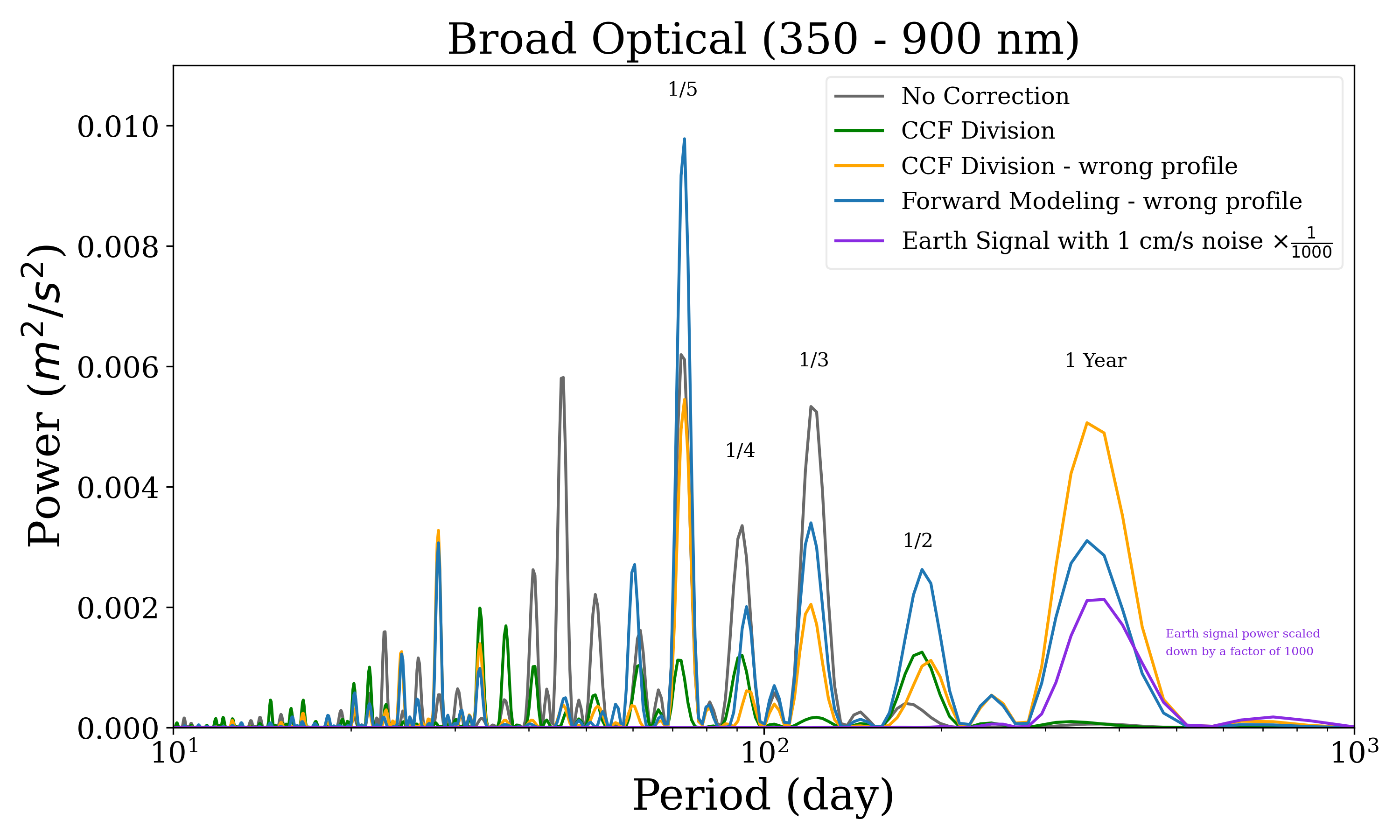}
\includegraphics[scale=0.55]{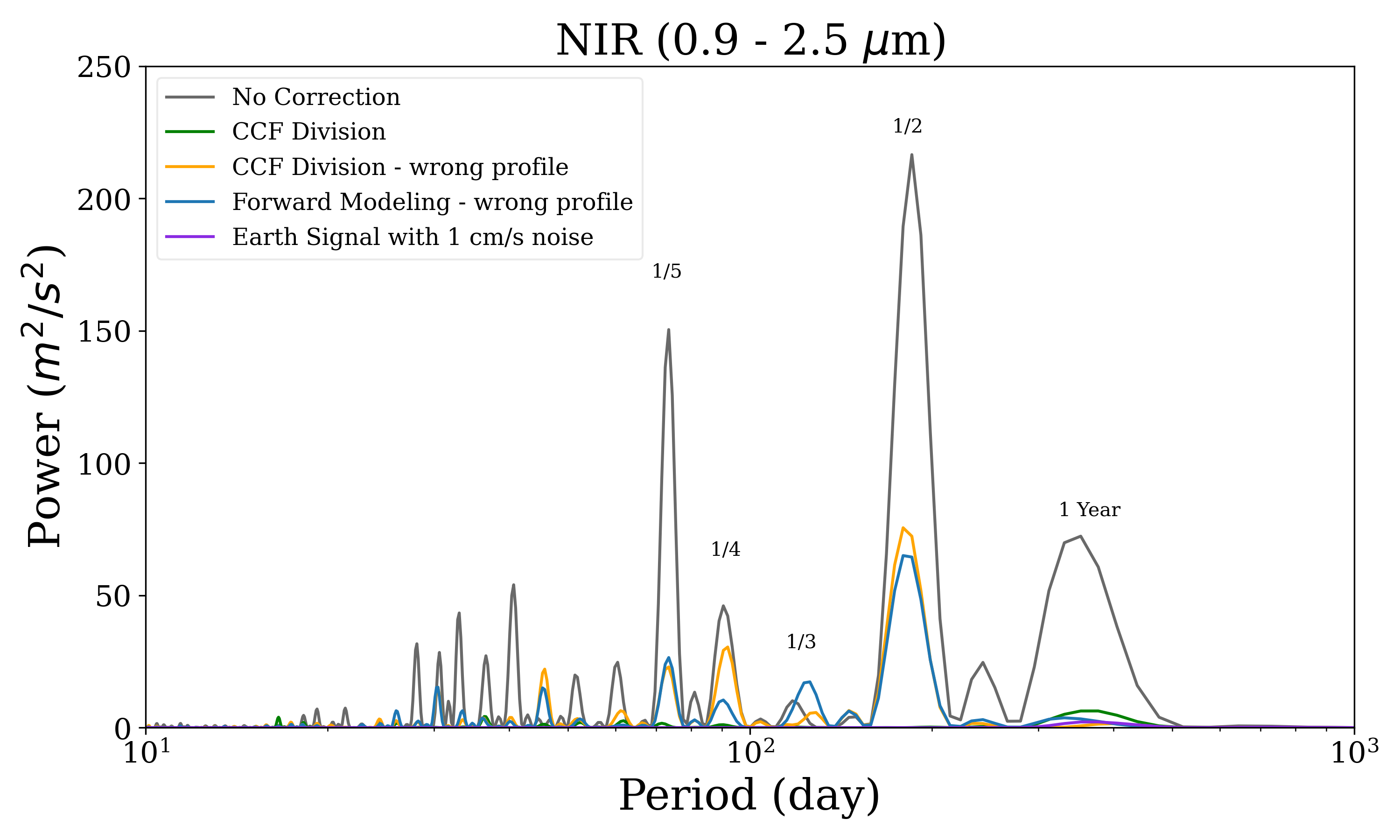}
\caption{Periodograms for the RVs plotted in Figure~\ref{fig:rvtime}, with the same RV data repeated for three years in the calculations of these periodograms to reduce the effects caused by the finite time baseline. The power in these Lomb-Scargle periodograms is unnormalized with the unit of (m/s)$^2$. The period of one year and some of its harmonics are labeled on top of the corresponding peaks. The power for the Earth's signal was computed with 1 cm/s white noise added, and in the top plot it is scaled down by a factor of 1000 to show its position, while it is barely visible in the bottom plot, highlighting the challenge in the NIR. See Section~\ref{sec:timeseries} for more detail.
\label{fig:periodograms}}
\end{figure}

Similar to how telluric induced RV errors manifest as coherent systematic patterns in the RV-BC plane, these errors appear as correlated noise with structures in the final RV time series as well, as shown in Figure~\ref{fig:rvtime}. Similar to the bottom panels of Figure~\ref{fig:pabcalc}, the characteristic width of the systematic patterns are determined by the typical line width. In addition, the patterns in Figure~\ref{fig:rvtime} are symmetric against the median date of the year due to the fact that we have set the BC values symmetric, following a $\tau$ Ceti style BC pattern but simplified with a simple sinusoidal curve (see Section~\ref{sec:method}). Notably, in the optical, the amplitudes of the telluric-induced RV errors are small compared to the RV signals of an Earth-like planet (the top panel of Figure~\ref{fig:rvtime}), even when tellurics are completely being ignored and left untreated (largely thanks to the weighting process which eliminated the bad regions). In contrast, tellurics could pose significant challenges to the detection of an Earth-like planet in the NIR (the bottom panel), creating systematic noise much larger than 10~cm/s even with the typical best-effort treatment (CCF division). These coherent systematic noises would add strong features to the periodograms for the NIR RVs at a periodicity of one year and its harmonics, while almost negligible in the optical, as illustrated in Figure~\ref{fig:periodograms}.

\section{Discussion and Conclusion}\label{sec:discussion}

\subsection{Caveats and Additional Limitations Set By Tellurics}

We enumerate the caveats of this study below, including several additional RV error terms that are induced by the telluric contamination but not considered in our simulations above. Because of these caveats and limitations, the numbers presented in this work should not be taken as the hard limit for the RV precision ceiling set by tellurics, but rather, an estimate of where the ceiling \textit{could} hit in real work. Improvements in instruments and algorithms could bring this ceiling down, but complications in real observations (such as the emission lines) could bring the ceiling up again. 

First, we assumed a very idealized level of precision, accuracy, and our knowledge on the extracted, observed spectra. We did not include photon noise in our simulated spectra in order to isolate the effects of telluric contamination. In reality, a limited SNR would translate into a limited ability of dividing out or modeling the telluric lines, and thus the existence of photon noise could further hinder our ability of eliminating the effects of tellurics. In addition, our simulation often assumes perfect knowledge of the spectral continuum or the blaze function, which is unrealistic for most RV instruments. The exact shape of the overall blaze function could change due to a varieties of factors, such as changes in the instruments and in the incident light (e.g., due to the time-dependent Rayleigh scattering; even changes in the polarization, e.g., \citealt{halverson2015}). Our work does not take such effects into accounts as this is beyond the scope of this study, but we do see some of the effects of a changing continuum in the blue region: in the calculations with the Forward Modeling using the wrong telluric profiles, we did not correct for the continuum absorption caused by the ozone in the blue optical region, which caused additional RV errors compared with the CCF division method which corrects for this continuum change (see the bottom panel of Figure~\ref{fig:rmswave}). 

We also assumed perfect knowledge of the true underlying stellar spectrum, i.e., having a perfect stellar template or mask. In realistic situations, stellar spectra of very high SNR can be recovered from stacking many observations, but telluric lines cannot be completely eliminated from the stacked spectrum and spectral normalization can be challenging. Residual tellurics in the stellar template could induce additional errors as they would ``latch" onto the telluric lines in the observed spectrum at each epoch (e.g., \citealt{wang2019}). Therefore, constructing a clear stellar template is vital to minimizing the impact of tellurics in the RV but can be a challenging task, especially for the NIR. In particular, it is extra challenging to separate the telluric lines from stellar lines for targets near the ecliptic poles that would have a small BC variation (and unfortunately, such targets are very often within the continuous viewing zones of space missions such as TESS and JWST). Future work on constructing accurate stellar templates would no doubt bring significant improvement in the RV precision, and it is also potentially critical for mitigating stellar jitter by anchoring an accurate ``median'' among the varying stellar spectral time series.

Furthermore, we adopted a very simple scenario of telluric contamination and the level of incompetency in modeling the tellurics. First, we created a simple imitation to capture the mismatch between the real telluric lines and the model lines (see the description about the Forward Modeling method in Section~\ref{sec:method}), while in reality the discrepancy could be larger. We have not tested the effects at various degrees or types of mismatches. Second, we did not include sky emission lines, such as the OH and Na lines, which could be challenging to remove due to their high variability and high intensity, especially for fiber-fed spectrographs without efficient sky fibers and particularly challenging in the NIR. Better algorithms in mitigating our lack of knowledge in the varying telluric absorption and emission lines is one of the critical next steps in this line of work, and we discuss some promising future leads in the next subsection.

There are also some caveats in the algorithms for extracting RVs in our work, especially when fitting the tellurics. Modeling the tellurics, the stellar spectrum, and the RV simultaneously is a non-linear optimization problem, and we have not tested our algorithm extensively beyond comparing a few options and making a judgement call on choosing the ``apparent best one" in terms of the level of goodness of fit and stability in convergence under different initial conditions. We chose to ignore an additional correction on the continuum for simplicity, but such an additional term could fix the errors due to our omission of ozone absorption in the blue and further improve the RV precision (though complicates the algorithm in terms of better convergence). We also chose to model the tellurics separately in each spectral order, instead of performing a global fit of all orders, and we also did not fit a certain molecular species if it does not have enough absorption as quantified in its level of Doppler content. This is to follow the common practice in extracting RVs and also to easily isolate the effects of tellurics across different wavelengths, but it is unclear whether this is the best practice for obtaining a better fit to the tellurics and a higher RV precision. A more advanced algorithm for extracting RVs in the context of residual telluric lines is another key step in making progress in mitigating tellurics, or any form of spectral contamination.

We have limited the scope of this paper to working with spectra with a resolution of $R=120,000$, which is the common value for the current state-of-the-art RV spectrographs. However, a higher spectral resolution could bring additional benefits in terms of mitigating the tellurics (certainly for the intrinsic Doppler precision and mitigating stellar jitter; e.g., \citealt{davis2017}). For example, the ``convolution" errors caused by division shown in Figure~\ref{fig:conv} would decrease as the spectral resolution increases and the stellar and telluric lines are relatively more resolved from each other. A higher resolution might also help with constructing a more accurate telluric model and thus reducing the errors caused by an incomplete knowledge of the telluric lines, such as the ones we see with the ``K-profile" methods in this work. However, the resolution adopted in this work (R = 120,000) is already high enough to resolve most of the telluric lines very well. However, for some of the intrinsically heavily blended lines, such as the water lines near 1.5~$\mu$m, a higher resolution up to R $\sim$ 200,000 does seem to bring a visible though limited improvement in resolving them. 

Finally, all calculations and discussions in this work are limited to solar analogs. RV measurements on stars of different stellar types will differ, as the Doppler information content of each star distributes differently across the wavelengths, and they also interact differently with the telluric lines, as a result. Probing how the effects of tellurics change with different stellar types is also very important, especially for the NIR spectrographs that primarily target M dwarfs. However, this is beyond the scope of this paper and will be the focus of a follow-up paper of this work.

\subsection{Conclusion and Future Work}

In this work, we characterized the effects of telluric absorption in the broad optical and NIR bands using simulations, and we compared the efficacy of two commonly adopted methods for mitigating tellurics under different assumptions, including simulating the lack of knowledge on the telluric line profile. We found that tellurics contribute to the RV error budget on the level of a few cm/s in the optical regime, but this could increase to m/s level in the NIR, depending on the exact wavelength range. In addition to adding scatter to the RVs, tellurics induce coherent systematic signals, which could be more problematic for the search of periodic planetary signals.

We found that, under realistic assumptions where the telluric lines are not perfectly known, both methods, CCF$+$division and Forward Modeling, perform reasonably well in mitigating tellurics, especially in the NIR. In the optical bands, it is important and effective to simply lower the weights of the spectral regions that are affected by tellurics, or eliminate them completely. One could easily argue that the weighting scheme adopted in this work is not the optimized one, for example, as discussed in Section~\ref{subsec:errorbands} when comparing the combined RVs from the CCF and Forward Modeling methods using wrong profiles. With an appropriate weighting or elimination scheme, it would not be necessary to invoke sophisticated mitigating methods such as the ones presented in this paper to reach a 10~cm/s precision in the broad optical region.

In addition, it is more important to have an accurate model for the telluric lines in the observed spectrum. Under the assumptions in this work where we adopted a set of wrong line profiles for the tellurics (inducing a residual error on the level of a few percent), the RV errors induced by the tellurics show a significant increase, especially in the NIR. This suggests that modeling the telluric lines more accurately, e.g., to better than a couple of percent, should bring visible improvement in the RV precision. 

The purpose of this work is to assess the impacts of tellurics and some of the commonly adopted mitigation strategies. There are certainly works to be done to explore better strategies in mitigating tellurics. For example, as mentioned above, perhaps a higher spectral resolution would bring an improvement in disentangling the tellurics, and there might be an optimal spectral resolution beyond which the gain in telluric mitigation would no longer be significant. 

From the software side, there is still plenty of room for improvement: algorithms for constructing the most accurate template from observations in the existence of tellurics, algorithms for modeling the telluric lines more accurately, and also better algorithms to solve for the RVs in the telluric-contaminated spectra. For example, the RV extraction code \texttt{wobble} by \cite{bedell2019}, shows an innovative path towards solving the telluric problem. The algorithm adopted by \texttt{wobble} makes minimal assumptions regarding the prior knowledge on the stellar template or the telluric lines. Its pure machine-learning, data-driven approach offers an alternative path towards disentangling the stellar and telluric lines. This approach is similar to the PCA-based telluric modeling presented in \cite{artigau2014} and used in the CFHT-SPIRou pipeline (e.g., \citealt{donati2020,moutou2020}), but with a more data-driven way to construct a telluric-free stellar template. Although \texttt{wobble} suffers from ``convolution" errors as it assumes simple multiplication of spectra, it should, in principle, perform as well as the Division method presented in this work, which can reach or is close to the photon limited precision in the majority of the common photometric bands. \texttt{wobble} has been very successful in the optical band, and it is yet to see whether it could tackle the more challenging NIR region (e.g., \citealt{gan2021}). 

Another innovative data-driven method was presented by \cite{rajpaul2020}, where all stellar spectra were modeled with gaussian processes and RVs were measured through pairwise comparison, and telluric contamination was part of the ``variance" in the spectra. This has the advantage of not requiring any prior knowledge on tellurics and incorporating the impact of tellurics in the final error estimation, but similar to \texttt{wobble}, this algorithm is yet to be tested for the NIR. A plausible path might be to use the synthetic telluric models as inputs to help the data-driven algorithms to arrive at a more accurate solution for the tellurics (e.g., perhaps along the lines of the method developed by \citealt{cretignier2021}), while taking into account the uncertainties in telluric modeling, which would in turn generate a more accurate stellar model and more accurate RVs as a result. 

In addition, improvements can be made in weighing the spectral regions contaminated by tellurics. In this work, weights were only adopted on a per-order basis, but an order covers a relatively wide spectral range ($\sim$nm). Sub-order weights or even pixel-based weights or rejections have the potential to further improve the RV precision while only suffering a relatively small SNR loss, and there might be a balance between rejecting telluric pixels versus retaining SNR. 

Finally, it is noteworthy that in the NIR, tellurics could produce more errors than the stellar jitter. While stellar jitter remains the bottleneck for getting down to 10 cm/s in the optical (e.g., \citealt{eprv2016}), it is thought to be less of a problem in the NIR due to the decreased magnitude of spot contrast (e.g., \citealt{tran2021,cale2021}). However, the systematics induced by the tellurics in the NIR can get easily up to 1 m/s and above (e.g., Figure~\ref{fig:rvtime}), on par with the amplitudes of stellar jitter of mature and relatively quiet stars. Mitigating the effects of tellurics is thus vitally important for archiving sub-m/s RV precision in the NIR. If NIR proves to be critical in mitigating the stellar jitter to the $\sim$cm/s level, then achieving a $\sim$cm/s precision would become critical for the NIR. This means great efforts should be made to deal with the telluric contamination in the NIR, or launching a space-based NIR spectrograph, such as \textit{EarthFinder}, becomes very appealing.  

\acknowledgements

N. L. gratefully acknowledges support from an NSF GRFP. 
PPP acknowledge support from NASA (EarthFinder Probe Mission Concept Study 16-APROBES16-0020,  Exoplanet Research Program Award \#80NSSC20K0251, JPL Exoplanet Exploration Program and Research and Technology Development) and the NSF (Astronomy and Astrophysics Grants \#1716202 and 2006517).

This work is being supported by CNES (Centre National des Etudes Spatiales) and CNRS (Centre National de la Recherche Scientifique).
TAPAS is a service maintained by ETHER data center. ETHER acknowledges for TAPAS the use of HITRAN data base and the LBLRTM radiative transfer code, the use of ECMWF data and the ETHER data center. We acknowledge useful discussions with Iouli Gordon and Larry Rothman.

\bibliography{references}

\begin{thebibliography}{}
\expandafter\ifx\csname natexlab\endcsname\relax\def\natexlab#1{#1}\fi
\providecommand{\url}[1]{\href{#1}{#1}}

\bibitem[{{Artigau} {et~al.}(2014){Artigau}, {Astudillo-Defru}, {Delfosse},
  {Bouchy}, {Bonfils}, {Lovis}, {Pepe}, {Moutou}, {Donati}, {Doyon}, \&
  {Malo}}]{artigau2014}
{Artigau}, {\'E}., {Astudillo-Defru}, N., {Delfosse}, X., {et~al.} 2014, in
  Society of Photo-Optical Instrumentation Engineers (SPIE) Conference Series,
  Vol. 9149, Society of Photo-Optical Instrumentation Engineers (SPIE)
  Conference Series, 5

\bibitem[{{Artigau} {et~al.}(2021){Artigau}, {H{\'e}brard}, {Cadieux},
  {Vandal}, {Cook}, {Doyon}, {Gagn{\'e}}, {Moutou}, {Martioli}, {Frasca},
  {Jahandar}, {Lafreni{\`e}re}, {Malo}, {Donati}, {Cort{\'e}s-Zuleta},
  {Boisse}, {Delfosse}, {Carmona}, {Fouqu{\'e}}, {Morin}, {Rowe}, {Marino},
  {Papini}, {Ciardi}, {Lund}, {Martins}, {Pelletier}, {Arnold}, {Bouchy},
  {Forveille}, {Santos}, {Bonfils}, {Figueira}, {Fausnaugh}, {Ricker},
  {Latham}, {Seager}, {Winn}, {Jenkins}, {Ting}, {Torres}, \& {Gomes da
  Silva}}]{artigau2021}
{Artigau}, {\'E}., {H{\'e}brard}, G., {Cadieux}, C., {et~al.} 2021, \aj, 162,
  144

\bibitem[{{Baker} {et~al.}(2020){Baker}, {Blake}, \& {Reiners}}]{baker2020}
{Baker}, A.~D., {Blake}, C.~H., \& {Reiners}, A. 2020, \apjs, 247, 24

\bibitem[{{Baker} {et~al.}(2017){Baker}, {Blake}, \& {Sliski}}]{baker2017}
{Baker}, A.~D., {Blake}, C.~H., \& {Sliski}, D.~H. 2017, \pasp, 129, 085002

\bibitem[{{Baranne} {et~al.}(1996){Baranne}, {Queloz}, {Mayor}, {Adrianzyk},
  {Knispel}, {Kohler}, {Lacroix}, {Meunier}, {Rimbaud}, \& {Vin}}]{baranne1996}
{Baranne}, A., {Queloz}, D., {Mayor}, M., {et~al.} 1996, \aaps, 119, 373

\bibitem[{{Bean} {et~al.}(2010){Bean}, {Seifahrt}, {Hartman}, {Nilsson},
  {Wiedemann}, {Reiners}, {Dreizler}, \& {Henry}}]{bean2010}
{Bean}, J.~L., {Seifahrt}, A., {Hartman}, H., {et~al.} 2010, \apj, 713, 410

\bibitem[{{Bedell} {et~al.}(2019){Bedell}, {Hogg}, {Foreman-Mackey}, {Montet},
  \& {Luger}}]{bedell2019}
{Bedell}, M., {Hogg}, D.~W., {Foreman-Mackey}, D., {Montet}, B.~T., \& {Luger},
  R. 2019, \aj, 158, 164

\bibitem[{{Bertaux} {et~al.}(2014){Bertaux}, {Lallement}, {Ferron}, {Boonne},
  \& {Bodichon}}]{tapas}
{Bertaux}, J.~L., {Lallement}, R., {Ferron}, S., {Boonne}, C., \& {Bodichon},
  R. 2014, \aap, 564, A46

\bibitem[{{Butler} {et~al.}(1996){Butler}, {Marcy}, {Williams}, {McCarthy},
  {Dosanjh}, \& {Vogt}}]{butler1996}
{Butler}, R.~P., {Marcy}, G.~W., {Williams}, E., {et~al.} 1996, \pasp, 108, 500

\bibitem[{{Butler} {et~al.}(2017){Butler}, {Vogt}, {Laughlin}, {Burt},
  {Rivera}, {Tuomi}, {Teske}, {Arriagada}, {Diaz}, {Holden}, \&
  {Keiser}}]{butler2017}
{Butler}, R.~P., {Vogt}, S.~S., {Laughlin}, G., {et~al.} 2017, \aj, 153, 208

\bibitem[{{Cale} {et~al.}(2021){Cale}, {Reefe}, {Plavchan}, {Tanner}, {Gaidos},
  {Gagn{\'e}}, {Gao}, {Kane}, {B{\'e}jar}, {Lodieu}, {Anglada-Escud{\'e}},
  {Ribas}, {Pall{\'e}}, {Quirrenbach}, {Amado}, {Reiners}, {Caballero}, {Rosa
  Zapatero Osorio}, {Dreizler}, {Howard}, {Fulton}, {Xuesong Wang}, {Collins},
  {El Mufti}, {Wittrock}, {Gilbert}, {Barclay}, {Klein}, {Martioli},
  {Wittenmyer}, {Wright}, {Addison}, {Hirano}, {Tamura}, {Kotani}, {Narita},
  {Vermilion}, {Lee}, {Geneser}, {Teske}, {Quinn}, {Latham}, {Esquerdo},
  {Calkins}, {Berlind}, {Zohrabi}, {Stibbards}, {Kotnana}, {Jenkins},
  {Twicken}, {Henze}, {Kidwell}, {Burke},
  {Villase\{\textbackslash\raisebox{-0.5ex}\textasciitilde n\}or}, \&
  {Boyd}}]{cale2021}
{Cale}, B., {Reefe}, M., {Plavchan}, P., {et~al.} 2021, arXiv e-prints,
  arXiv:2109.13996

\bibitem[{{Crass} {et~al.}(2021){Crass}, {Gaudi}, {Leifer}, {Beichman},
  {Bender}, {Blackwood}, {Burt}, {Callas}, {Cegla}, {Diddams}, {Dumusque},
  {Eastman}, {Ford}, {Fulton}, {Gibson}, {Halverson}, {Haywood}, {Hearty},
  {Howard}, {Latham}, {L{\"o}hner-B{\"o}ttcher}, {Mamajek}, {Mortier},
  {Newman}, {Plavchan}, {Quirrenbach}, {Reiners}, {Robertson}, {Roy}, {Schwab},
  {Seifahrt}, {Szentgyorgyi}, {Terrien}, {Teske}, {Thompson}, \&
  {Vasisht}}]{crass2021}
{Crass}, J., {Gaudi}, B.~S., {Leifer}, S., {et~al.} 2021, arXiv e-prints,
  arXiv:2107.14291

\bibitem[{{Crepp} {et~al.}(2016){Crepp}, {Crass}, {King}, {Bechter}, {Bechter},
  {Ketterer}, {Reynolds}, {Hinz}, {Kopon}, {Cavalieri}, {Fantano}, {Koca},
  {Onuma}, {Stapelfeldt}, {Thomes}, {Wall}, {Macenka}, {McGuire}, {Korniski},
  {Zugby}, {Eisner}, {Gaudi}, {Hearty}, {Kratter}, {Kuchner}, {Micela},
  {Nelson}, {Pagano}, {Quirrenbach}, {Schwab}, {Skrutskie}, {Sozzetti},
  {Woodward}, \& {Zhao}}]{Crepp2016}
{Crepp}, J.~R., {Crass}, J., {King}, D., {et~al.} 2016, in Society of
  Photo-Optical Instrumentation Engineers (SPIE) Conference Series, Vol. 9908,
  Ground-based and Airborne Instrumentation for Astronomy VI, ed. C.~J.
  {Evans}, L.~{Simard}, \& H.~{Takami}, 990819

\bibitem[{{Cretignier} {et~al.}(2021){Cretignier}, {Dumusque}, {Hara}, \&
  {Pepe}}]{cretignier2021}
{Cretignier}, M., {Dumusque}, X., {Hara}, N.~C., \& {Pepe}, F. 2021, \aap, 653,
  A43

\bibitem[{{Cunha} {et~al.}(2014){Cunha}, {Santos}, {Figueira}, {Santerne},
  {Bertaux}, \& {Lovis}}]{cunha2014}
{Cunha}, D., {Santos}, N.~C., {Figueira}, P., {et~al.} 2014, \aap, 568, A35

\bibitem[{{Davis} {et~al.}(2017){Davis}, {Cisewski}, {Dumusque}, {Fischer}, \&
  {Ford}}]{davis2017}
{Davis}, A.~B., {Cisewski}, J., {Dumusque}, X., {Fischer}, D.~A., \& {Ford},
  E.~B. 2017, \apj, 846, 59

\bibitem[{{Donati} {et~al.}(2020){Donati}, {Kouach}, {Moutou}, {Doyon},
  {Delfosse}, {Artigau}, {Baratchart}, {Lacombe}, {Barrick}, {H{\'e}brard},
  {Bouchy}, {Saddlemyer}, {Par{\`e}s}, {Rabou}, {Micheau}, {Dolon}, {Reshetov},
  {Challita}, {Carmona}, {Striebig}, {Thibault}, {Martioli}, {Cook},
  {Fouqu{\'e}}, {Vermeulen}, {Wang}, {Arnold}, {Pepe}, {Boisse}, {Figueira},
  {Bouvier}, {Ray}, {Feugeade}, {Morin}, {Alencar}, {Hobson}, {Castilho},
  {Udry}, {Santos}, {Hernandez}, {Benedict}, {Vall{\'e}e}, {Gallou}, {Dupieux},
  {Larrieu}, {Perruchot}, {Sottile}, {Moreau}, {Usher}, {Baril}, {Wildi},
  {Chazelas}, {Malo}, {Bonfils}, {Loop}, {Kerley}, {Wevers}, {Dunn}, {Pazder},
  {Macdonald}, {Dubois}, {Carri{\'e}}, {Valentin}, {Henault}, {Yan}, \&
  {Steinmetz}}]{donati2020}
{Donati}, J.~F., {Kouach}, D., {Moutou}, C., {et~al.} 2020, \mnras, 498, 5684

\bibitem[{{Dressing} {et~al.}(2019){Dressing}, {Stark}, {Plavchan}, \&
  {Lopez}}]{dressing2019}
{Dressing}, C., {Stark}, C.~C., {Plavchan}, P., \& {Lopez}, E. 2019, \baas, 51,
  268

\bibitem[{{Figueira} {et~al.}(2012){Figueira}, {Kerber}, {Chacon}, {Lovis},
  {Santos}, {Lo Curto}, {Sarazin}, \& {Pepe}}]{figueira2012}
{Figueira}, P., {Kerber}, F., {Chacon}, A., {et~al.} 2012, \mnras, 420, 2874

\bibitem[{{Fischer} {et~al.}(2016){Fischer}, {Anglada-Escude}, {Arriagada},
  {Baluev}, {Bean}, {Bouchy}, {Buchhave}, {Carroll}, {Chakraborty}, {Crepp},
  {Dawson}, {Diddams}, {Dumusque}, {Eastman}, {Endl}, {Figueira}, {Ford},
  {Foreman-Mackey}, {Fournier}, {F{\H u}r{\'e}sz}, {Gaudi}, {Gregory},
  {Grundahl}, {Hatzes}, {H{\'e}brard}, {Herrero}, {Hogg}, {Howard}, {Johnson},
  {Jorden}, {Jurgenson}, {Latham}, {Laughlin}, {Loredo}, {Lovis}, {Mahadevan},
  {McCracken}, {Pepe}, {Perez}, {Phillips}, {Plavchan}, {Prato}, {Quirrenbach},
  {Reiners}, {Robertson}, {Santos}, {Sawyer}, {Segransan}, {Sozzetti},
  {Steinmetz}, {Szentgyorgyi}, {Udry}, {Valenti}, {Wang}, {Wittenmyer}, \&
  {Wright}}]{eprv2016}
{Fischer}, D.~A., {Anglada-Escude}, G., {Arriagada}, P., {et~al.} 2016, \pasp,
  128, 066001

\bibitem[{{Gan} {et~al.}(2021){Gan}, {Lin}, {Xuesong Wang}, {Mao},
  {Fouqu{\'e}}, {Stassun}, {Giacalone}, {Fukui}, {Murgas}, {Ciardi}, {Howell},
  {Collins}, {Shporer}, {Arnold}, {Barclay}, {Charbonneau}, {Christiansen},
  {Crossfield}, {Dressing}, {Elliott}, {Esparza-Borges}, {Evans}, {Gnilka},
  {Gonzales}, {Howard}, {Isogai}, {Kawauchi}, {Kurita}, {Liu}, {Livingston},
  {Matson}, {Narita}, {Palle}, {Parviainen}, {Rackham}, {Rodriguez}, {Rose},
  {Rudat}, {Schlieder}, {Scott}, {Vezie}, {Ricker}, {Vanderspek}, {Latham},
  {Seager}, {Winn}, \& {Jenkins}}]{gan2021}
{Gan}, T., {Lin}, Z., {Xuesong Wang}, S., {et~al.} 2021, arXiv e-prints,
  arXiv:2110.04220

\bibitem[{{Gordon} {et~al.}(2017){Gordon}, {Rothman}, {Hill}, {Kochanov},
  {Tan}, {Bernath}, {Birk}, {Boudon}, {Campargue}, {Chance}, {Drouin}, {Flaud},
  {Gamache}, {Hodges}, {Jacquemart}, {Perevalov}, {Perrin}, {Shine}, {Smith},
  {Tennyson}, {Toon}, {Tran}, {Tyuterev}, {Barbe}, {Cs{\'a}sz{\'a}r}, {Devi},
  {Furtenbacher}, {Harrison}, {Hartmann}, {Jolly}, {Johnson}, {Karman},
  {Kleiner}, {Kyuberis}, {Loos}, {Lyulin}, {Massie}, {Mikhailenko},
  {Moazzen-Ahmadi}, {M{\"u}ller}, {Naumenko}, {Nikitin}, {Polyansky}, {Rey},
  {Rotger}, {Sharpe}, {Sung}, {Starikova}, {Tashkun}, {Auwera}, {Wagner},
  {Wilzewski}, {Wcis{\l}o}, {Yu}, \& {Zak}}]{hitran2017}
{Gordon}, I.~E., {Rothman}, L.~S., {Hill}, C., {et~al.} 2017, \jqsrt, 203, 3

\bibitem[{{Gullikson} {et~al.}(2014){Gullikson}, {Dodson-Robinson}, \&
  {Kraus}}]{gullikson2014}
{Gullikson}, K., {Dodson-Robinson}, S., \& {Kraus}, A. 2014, \aj, 148, 53

\bibitem[{{Halverson} {et~al.}(2015){Halverson}, {Roy}, {Mahadevan}, \&
  {Schwab}}]{halverson2015}
{Halverson}, S., {Roy}, A., {Mahadevan}, S., \& {Schwab}, C. 2015, \apjl, 814,
  L22

\bibitem[{{Halverson} {et~al.}(2016){Halverson}, {Terrien}, {Mahadevan}, {Roy},
  {Bender}, {Stef{\'a}nsson}, {Monson}, {Levi}, {Hearty}, {Blake}, {McElwain},
  {Schwab}, {Ramsey}, {Wright}, {Wang}, {Gong}, \& {Roberston}}]{halverson2016}
{Halverson}, S., {Terrien}, R., {Mahadevan}, S., {et~al.} 2016, in Society of
  Photo-Optical Instrumentation Engineers (SPIE) Conference Series, Vol. 9908,
  Ground-based and Airborne Instrumentation for Astronomy VI, ed. C.~J.
  {Evans}, L.~{Simard}, \& H.~{Takami}, 99086P

\bibitem[{{Howard} \& {Fulton}(2016)}]{howard2016}
{Howard}, A.~W., \& {Fulton}, B.~J. 2016, \pasp, 128, 114401

\bibitem[{{Kurucz}(2005)}]{kurucz2005}
{Kurucz}, R.~L. 2005, Memorie della Societa Astronomica Italiana Supplementi,
  8, 14

\bibitem[{{Li} {et~al.}(2018){Li}, {Blake}, {Nidever}, \& {Halverson}}]{li2018}
{Li}, D., {Blake}, C.~H., {Nidever}, D., \& {Halverson}, S.~P. 2018, \pasp,
  130, 014501

\bibitem[{{Lisogorskyi} {et~al.}(2019){Lisogorskyi}, {Jones}, \&
  {Feng}}]{lisogorskyi2019}
{Lisogorskyi}, M., {Jones}, H.~R.~A., \& {Feng}, F. 2019, \mnras, 485, 4804

\bibitem[{{Meier Vald{\'e}s} {et~al.}(2021){Meier Vald{\'e}s}, {Morris}, \&
  {Demory}}]{meier2021}
{Meier Vald{\'e}s}, E.~A., {Morris}, B.~M., \& {Demory}, B.~O. 2021, \aap, 649,
  A132

\bibitem[{{Morgan} {et~al.}(2021){Morgan}, {Savransky}, {Turmon}, {Mennesson},
  {Dula}, {Keithly}, {Mamajek}, {Newman}, {Plavchan}, {Robinson}, {Roudier}, \&
  {Stark}}]{morgan2021}
{Morgan}, R., {Savransky}, D., {Turmon}, M., {et~al.} 2021, Journal of
  Astronomical Telescopes, Instruments, and Systems, 7, 021220

\bibitem[{{Moutou} {et~al.}(2020){Moutou}, {Dalal}, {Donati}, {Martioli},
  {Folsom}, {Artigau}, {Boisse}, {Bouchy}, {Carmona}, {Cook}, {Delfosse},
  {Doyon}, {Fouqu{\'e}}, {Gaisn{\'e}}, {H{\'e}brard}, {Hobson}, {Klein},
  {Lecavelier des Etangs}, \& {Morin}}]{moutou2020}
{Moutou}, C., {Dalal}, S., {Donati}, J.~F., {et~al.} 2020, \aap, 642, A72

\bibitem[{{Plavchan} {et~al.}(2015){Plavchan}, {Latham}, {Gaudi}, {Crepp},
  {Dumusque}, {Furesz}, {Vanderburg}, {Blake}, {Fischer}, {Prato}, {White},
  {Makarov}, {Marcy}, {Stapelfeldt}, {Haywood}, {Collier-Cameron},
  {Quirrenbach}, {Mahadevan}, {Anglada}, \& {Muirhead}}]{exopag2015}
{Plavchan}, P., {Latham}, D., {Gaudi}, S., {et~al.} 2015, ArXiv e-prints,
  arXiv:1503.01770

\bibitem[{{Plavchan} {et~al.}(2020){Plavchan}, {Vasisht}, {Beichman}, {Cegla},
  {Dumusque}, {Wang}, {Gao}, {Dressing}, {Bastien}, {Basu}, {Beatty},
  {Bechter}, {Bechter}, {Blake}, {Bourrier}, {Cale}, {Ciardi}, {Crass},
  {Crepp}, {de Kleer}, {Diddams}, {Eastman}, {Fischer}, {Gagn{\'e}}, {Gaudi},
  {Grier}, {Hall}, {Halverson}, {Hamze}, {Herrero Casas}, {Howard}, {Kempton},
  {Latouf}, {Leifer}, {Lightsey}, {Lisse}, {Martin}, {Matzko}, {Mawet}, {Mayo},
  {Newman}, {Papp}, {Pope}, {Purcell}, {Quinn}, {Ribas}, {Rosich},
  {Sanchez-Maes}, {Tanner}, {Thompson}, {Vahala}, {Wang}, {Williams}, {Wise},
  \& {Wright}}]{plavchan2020}
{Plavchan}, P., {Vasisht}, G., {Beichman}, C., {et~al.} 2020, arXiv e-prints,
  arXiv:2006.13428

\bibitem[{{Rajpaul} {et~al.}(2020){Rajpaul}, {Aigrain}, \&
  {Buchhave}}]{rajpaul2020}
{Rajpaul}, V.~M., {Aigrain}, S., \& {Buchhave}, L.~A. 2020, \mnras, 492, 3960

\bibitem[{{Rothman} {et~al.}(2013){Rothman}, {Gordon}, {Babikov}, {Barbe},
  {Chris Benner}, {Bernath}, {Birk}, {Bizzocchi}, {Boudon}, {Brown},
  {Campargue}, {Chance}, {Cohen}, {Coudert}, {Devi}, {Drouin}, {Fayt}, {Flaud},
  {Gamache}, {Harrison}, {Hartmann}, {Hill}, {Hodges}, {Jacquemart}, {Jolly},
  {Lamouroux}, {Le Roy}, {Li}, {Long}, {Lyulin}, {Mackie}, {Massie},
  {Mikhailenko}, {M{\"u}ller}, {Naumenko}, {Nikitin}, {Orphal}, {Perevalov},
  {Perrin}, {Polovtseva}, {Richard}, {Smith}, {Starikova}, {Sung}, {Tashkun},
  {Tennyson}, {Toon}, {Tyuterev}, \& {Wagner}}]{hitran2013}
{Rothman}, L.~S., {Gordon}, I.~E., {Babikov}, Y., {et~al.} 2013, \jqsrt, 130, 4

\bibitem[{{Roy} {et~al.}(2020){Roy}, {Halverson}, {Mahadevan}, {Stefansson},
  {Monson}, {Logsdon}, {Bender}, {Blake}, {Golub}, {Gupta}, {Jaehnig},
  {Kanodia}, {Kaplan}, {McElwain}, {Ninan}, {Rajagopal}, {Robertson}, {Schwab},
  {Terrien}, {Wang}, {Wolf}, \& {Wright}}]{roy2020}
{Roy}, A., {Halverson}, S., {Mahadevan}, S., {et~al.} 2020, \aj, 159, 161

\bibitem[{{Sameshima} {et~al.}(2018){Sameshima}, {Matsunaga}, {Kobayashi},
  {Kawakita}, {Hamano}, {Ikeda}, {Kondo}, {Fukue}, {Taniguchi}, {Mizumoto},
  {Arai}, {Otsubo}, {Takenaka}, {Watase}, {Asano}, {Yasui}, {Izumi}, \&
  {Yoshikawa}}]{sameshima2018}
{Sameshima}, H., {Matsunaga}, N., {Kobayashi}, N., {et~al.} 2018, \pasp, 130,
  074502

\bibitem[{{Seifahrt} {et~al.}(2010){Seifahrt}, {K{\"a}ufl}, {Z{\"a}ngl},
  {Bean}, {Richter}, \& {Siebenmorgen}}]{seifahrt2010}
{Seifahrt}, A., {K{\"a}ufl}, H.~U., {Z{\"a}ngl}, G., {et~al.} 2010, \aap, 524,
  A11

\bibitem[{{Sithajan} {et~al.}(2016){Sithajan}, {Ge}, \& {Wang}}]{sithajan2016}
{Sithajan}, S., {Ge}, J., \& {Wang}, J. 2016, in American Astronomical Society
  Meeting Abstracts, Vol. 227, American Astronomical Society Meeting Abstracts,
  137.19

\bibitem[{{Smette} {et~al.}(2015){Smette}, {Sana}, {Noll}, {Horst}, {Kausch},
  {Kimeswenger}, {Barden}, {Szyszka}, {Jones}, {Gallenne}, {Vinther},
  {Ballester}, \& {Taylor}}]{smette2015}
{Smette}, A., {Sana}, H., {Noll}, S., {et~al.} 2015, \aap, 576, A77

\bibitem[{{Tran} {et~al.}(2021){Tran}, {Bowler}, {Cochran}, {Endl},
  {Stefansson}, {Mahadevan}, {Ninan}, {Bender}, {Halverson}, {Roy}, \&
  {Terrien}}]{tran2021}
{Tran}, Q.~H., {Bowler}, B.~P., {Cochran}, W.~D., {et~al.} 2021, arXiv
  e-prints, arXiv:2101.11005

\bibitem[{{Ulmer-Moll} {et~al.}(2019){Ulmer-Moll}, {Figueira}, {Neal},
  {Santos}, \& {Bonnefoy}}]{ulmer-moll2019}
{Ulmer-Moll}, S., {Figueira}, P., {Neal}, J.~J., {Santos}, N.~C., \&
  {Bonnefoy}, M. 2019, \aap, 621, A79

\bibitem[{{Vacca} {et~al.}(2003){Vacca}, {Cushing}, \& {Rayner}}]{vacca2003}
{Vacca}, W.~D., {Cushing}, M.~C., \& {Rayner}, J.~T. 2003, \pasp, 115, 389

\bibitem[{{Wright} \& {Robertson}(2017)}]{eprv3}
{Wright}, J.~T., \& {Robertson}, P. 2017, Research Notes of the American
  Astronomical Society, 1, 51

\bibitem[{{Wright} {et~al.}(2013){Wright}, {Roy}, {Mahadevan}, {Wang}, {Ford},
  {Payne}, {Lee}, {Wang}, {Crepp}, {Gaudi}, {Eastman}, {Pepper}, {Ge},
  {Fleming}, {Ghezzi}, {Gonz{\'a}lez-Hern{\'a}ndez}, {Cargile}, {Stassun},
  {Wisniewski}, {Dutra-Ferreira}, {Porto de Mello}, {Maia}, {Nicolaci da
  Costa}, {Ogando}, {Santiago}, {Schneider}, \& {Hearty}}]{wright2013}
{Wright}, J.~T., {Roy}, A., {Mahadevan}, S., {et~al.} 2013, \apj, 770, 119

\bibitem[{{Xuesong Wang} {et~al.}(2019){Xuesong Wang}, {Wright}, {Bender},
  {Howard}, {Isaacson}, {Veyette}, \& {Muirhead}}]{wang2019}
{Xuesong Wang}, S., {Wright}, J.~T., {Bender}, C., {et~al.} 2019, \aj, 158, 216

\bibitem[{{Zechmeister} {et~al.}(2018){Zechmeister}, {Reiners}, {Amado},
  {Azzaro}, {Bauer}, {B{\'e}jar}, {Caballero}, {Guenther}, {Hagen}, {Jeffers},
  {Kaminski}, {K{\"u}rster}, {Launhardt}, {Montes}, {Morales}, {Quirrenbach},
  {Reffert}, {Ribas}, {Seifert}, {Tal-Or}, \& {Wolthoff}}]{zeichmeister2018}
{Zechmeister}, M., {Reiners}, A., {Amado}, P.~J., {et~al.} 2018, \aap, 609, A12

\end{thebibliography}
\end{CJK*}

\end{document}